\documentclass[preprintnumbers,secnumarabic,amsmath,amssymb,nofootinbib,floatfix]{revtex4}
\usepackage{varioref,exscale,latexsym,amsmath,amssymb}
\usepackage{graphicx,epstopdf}
\usepackage{epsfig}
\usepackage{graphicx}% Include figure files
\usepackage{dcolumn}% Align table columns on decimal point
\usepackage{bm}% bold math
\usepackage{simplewick}

\def\beq{\begin{equation}}

\def\eeq{\end{equation}}

\def\beqa{\begin{eqnarray}}

\def\eeqa{\end{eqnarray}}

\begin{document}

\title{{\bf Non-local quantum effects in cosmology 1: Quantum memory, non-local FLRW equations and singularity avoidance }}

\medskip\
{\author{ John F. Donoghue}
%\email[Email: ]{donoghue@physics.umass.edu}
\author{Basem Kamal El-Menoufi}
%\email[Email: ]{bmahmoud@physics.umass.edu}
\affiliation{Department of Physics,
University of Massachusetts\\
Amherst, MA  01003, USA}

\begin{abstract}
We discuss cosmological effects of the quantum loops of massless particles, which lead to temporal non-localities in the equations of motion governing the scale factor $a(t)$. For the effects discussed here, loops cause the evolution of $a(t)$ to depend on the memory of the curvature in the past with a weight that scales initially as $1/(t-t')$. As one of our primary examples we discuss the situation with a large number of light particles, such that these effects occur in a region where gravity may still be treated classically. However, we also describe the effect of quantum graviton loops and the full set of Standard Model particles. We show that these effects decrease with time in an expanding phase, leading to classical behavior at late time. In a contracting phase, within our approximations the quantum results can lead to a bounce-like behavior at scales below the Planck mass, avoiding the singularities required classically by the Hawking-Penrose theorems. For conformally invariant fields, such as the Standard Model with a conformally coupled Higgs, this result is purely non-local and parameter independent.
\end{abstract}
%\vspace{0.2 in}
%\end{titlepage}
%\setcounter{page}{0}
%\newpage
\maketitle
\section{Introduction}

Massless particles can propagate over long distances, and loops involving massless particles generate nonlocal effects. In cosmology, where
the evolution of the scale factor depends only on time, this means that loops can generate temporal non-localities. There will be modifications
to the FLRW (Friedmann, Lema\^{i}tre, Robertson, Walker) equations governing the scale factor $a(t)$, which in the classical theory are local differential equations. The effects of loops will generate new contributions where the equation for the scale factor depends on what the scale factor was doing in the past.
We refer to this effect as the quantum memory of the scale factor and it is the subject of the present paper.
Such non-local effects are calculable, even if we do not know the full theory of quantum gravity, because they come from the low energy portion of the effective field theory \cite{eft} where the interactions are those of general relativity.

Quantum non-local effects produce modifications to standard cosmological behavior at scales below, but approaching, the Planck scale. In an expanding universe, we explore how classical behavior emerges from the quantum regime. In a contracting universe, singularities are inevitable in the classical theory, as
shown by the Hawking-Penrose singularity theorems \cite{singularity}. We study whether quantum effects could lead to the avoidance of
singularities. Our work contains some approximations, described below, but within the context of those approximations it does seem that quantum effects do lead to non-singular bounce solutions in at least some situations.

We will provide results for all forms of matter. However, two cases are of particular importance. One is obviously pure gravity, studying the effects of
graviton loops. The other is the case of a large number of matter fields. Conceptually this situation is distinctive
because when the number ($N$) of matter fields is large,
the non-local quantum effects become important at a scale $M_P/\sqrt{N}$, at which point general relativity itself can be treated classically. For example,
in such a theory the effect of the graviton vacuum polarization from $N$ scalar particles diagram of Fig. 1 can be summed to produce a modification to the graviton propagator
\begin{equation}
\frac{1}{q^2} \to \frac{1}{q^2 - \frac{G N q^2}{120\pi} \log (-q^2/\mu^2)}\, .
\end{equation}
The logarithmic term is crucial for restoring unitarity to scattering amplitudes in this theory \cite{han, self}. It is the momentum space equivalent of the non-local terms that we will be studying in this paper. We are interested in the effect of this loop, not in scattering amplitudes but in cosmology. The large $N$ limit is also relevant for the physical universe, as the Standard Model has roughly a hundred effective degrees of freedom (fermions, vectors and scalars, as defined in Sec. 4) producing quantum effects that are larger than graviton loops. We also display results for the Standard Model set of particles.

%%%%%%%%%%%%%%%%%%%%%%%%%
\begin{figure}[ht]
\centerline{
\includegraphics[width=.25\textwidth]{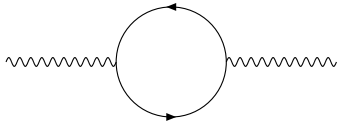}
}
\caption{Vacuum polarization graph. }
\label{VP}
\end{figure}
%%%%%%%%%%%%%%%%%%%%%%%%%%%%%%

The study of quantum field theory and gravity is a vast subject - many fundamental developments can be traced in the references of books such as \cite{Birrell:1982ix, parkertoms, bos}. In connection with non-localities, we should mention some previous work in particular. Barvinsky, Vilkovisky and collaborators \cite{Barvinsky:1985an, Barvinsky:1995jv, Barvinsky:1994ic, Avramidi:1990je, gospel}
have developed powerful heat kernel techniques to uncover non-local effects. We use some of their results in Sec. 4. Espriu and collaborators \cite{Espriu:2005qn, Cabrer:2007xm} have made
important preliminary investigations of possible non-local effects during inflation. We are building on these earlier results. In addition there are a
wide variety of works in non-local models (see for example \cite{various} and references therein) which however are of a quite different character than the
quantum effects that we study. 

The plan of the paper is as follows. In Sec. 2 and 3, we first treat simple perturbation theory around flat space. This is useful to show the nature of the non-locality in time, and to show how one obtains causal behavior in the equations of motion. We then provide a non-linear form of this result, matching to
the heat kernal methods in Sec. 4, with the corresponding non-linear FLRW equations of motion being displayed in Sec. 5. The expanding universe emerging from the quantum regime is studied in Sec. 6, while Sec. 7 is devoted to the exploration of singularity avoidance in a collapsing phase. Comments, caveats and further work are discussed in the summary.

\section{Perturbative analysis}

We first start with a perturbative treatment of the graviton vacuum polarization. This provides us with a basis for later treatment of the non-linear equations, separating the non-local effect from the renormalization of the local terms in the action. It also allows us to explore the impact of using the appropriate field theoretic formalism to generate causal behavior for cosmology in the next section.

We compute perturbatively the effective action for a massless free scalar field minimally coupled to gravity with the Lagrangian
\begin{align}\label{lag}
\mathcal{L} = \frac{1}{2} \sqrt{g} g^{\mu\nu} \partial_{\mu} \phi \partial_{\nu} \phi\, .
\end{align}
After performing the functional integral, the operator of interest reads
\begin{align}
\nonumber
\mathcal{D}&=\sqrt{g} (\Box) \\
&=\sqrt{g}g^{\mu\nu}\left(\partial_{\mu}\partial_{\nu}-\Gamma^{\alpha}_{\mu\nu}\partial_{\alpha}\right)\, .
\end{align}
\noindent The last equality holds because the covariant d'Alembertian acts on a scalar field. The metric is expanded around flat space (we use the mostly minus signature)
\begin{align}\label{metricsplit}
g_{\mu\nu} = \eta_{\mu\nu} + h_{\mu\nu}\, .
\end{align}
Likewise, the differential operator can be expanded in powers of $h_{\mu\nu}$ to yield
\begin{align}
\mathcal{D}= \partial^2 + \delta^{(1)} + \delta^{(2)} + \mathcal{O}(h^3)
\end{align}
where,
\begin{eqnarray}
&\partial^2 = \eta^{\mu\nu} \partial_{\mu} \partial_{\nu}, \quad \delta^{(1)} = -h^{\mu\nu}\partial_{\mu}\partial_{\nu} + \frac{1}{2} h \partial^2 - \eta^{\mu\nu} \underline{\Gamma}^{\alpha}_{\mu\nu}\partial_{\alpha}\\
&\delta^{(2)} = h^{\mu \nu} h_{\nu}^{\alpha} \partial_{\mu} \partial_{\alpha} - \frac{1}{2} h h^{\mu \nu}\partial_{\mu} \partial_{\nu} + \left(\frac{1}{4}h_{\mu \nu} h^{\mu \nu} + \frac{1}{8} h^2 \right) \partial^2
+ \left(h^{\mu\nu} + \frac{1}{2} h \eta^{\mu\nu} \right) \underline{\Gamma}^{\alpha}_{\mu\nu} \partial_{\partial} - \eta^{\mu\nu} \underline{\underline{\Gamma}}^{\alpha}_{\mu\nu} \partial_{\alpha} \, .
\end{eqnarray}
The indices are raised and lowered using the flat metric, and we have defined
\begin{align}
\underline{\Gamma}^{\alpha}_{\mu\nu} &= \frac{1}{2} \left(\partial_{\mu} h _{\nu}^{\alpha} + \partial_{\nu} h_{\mu}^{\alpha} - \partial^{\alpha} h_{\mu\nu} \right)\\
\underline{\underline{\Gamma}}^{\alpha}_{\mu\nu} &= -\frac{1}{2} h^{\alpha\beta} \left(\partial_{\mu} h_{\nu\beta} + \partial_{\nu} h_{\mu\beta} - \partial_{\beta} h_{\mu\nu} \right)\, .
\end{align}

To find the effective action, we take the logarithm of the differential operator and expand in powers of $h_{\mu\nu}$ to find
\begin{align}
Tr(\log \mathcal{D}) &= Tr (\log \partial^2) + Tr \left(G \delta^{(1)} + G \delta^{(2)} - \frac{1}{2} G \delta^{(1)} G \delta^{(1)}\right) + \mathcal{O}(h^3)\, .
\end{align}

\noindent In the above, $G$ is the Feynmann propagator of a massless scalar. Terms with one propagator vanish when regularized dimensionally. The first non vanishing contribution is at second order in $h_{\mu\nu}$.
We find at this order
\begin{align}
Tr(\log \mathcal{D})=-\frac{1}{2}& \int \frac{d^4k}{(2\pi)^4} h^{\mu\nu}(k)h^{\alpha\beta}(-k)\int \frac{d^4p}{(2\pi)^4}\frac{V_{\mu\nu}(k,p) V_{\alpha\beta}(k,p)}{(p^2 + i0)((p+k)^2 + i0)}
\end{align}
where
\begin{align}
V_{\mu\nu}(k,p) = p_{\mu} p_{\nu} - \frac{1}{2} \eta_{\mu\nu} p^2 + \frac{1}{2} k_{\mu} p_{\nu} + \frac{1}{2} k_{\nu} p_{\mu} - \frac{1}{2} \eta_{\mu\nu} k \cdot p \, .
\end{align}
This can be calculated straightforwardly, with the final result
\begin{equation}
Tr(\log \mathcal{D})=-\frac{1}{2} \int \frac{d^4k}{(2\pi)^4} h^{\mu\nu}(k)h^{\alpha\beta}(-k)T_{\mu\nu\alpha\beta}(k)
\end{equation}
where
\begin{align}
\nonumber
T_{\mu \nu \alpha \beta}(k) = \frac{i}{3840 \pi^2} \left(\frac{1}{\bar{\epsilon}} - \log \left(\frac{-k^2}{\mu^2}\right) \right) [&k^4 \left(6 \eta_{\mu \nu} \eta_{\alpha \beta} + \eta_{\mu \alpha}\eta_{\nu \beta} + \eta_{\mu \beta}\eta_{\nu \alpha}\right) + 8 k_{\mu} k_{\nu} k_{\alpha} k_{\beta} \\
- &k^2 \left(6 k_{\mu} k_{\nu} \eta_{\alpha \beta} + 6 k_{\alpha} k_{\beta} \eta_{\mu \nu} + k_{\mu} k_{\alpha} \eta_{\nu \beta} + k_{\mu} k_{\beta} \eta_{\nu \alpha} + k_{\nu} k_{\alpha} \eta_{\mu \beta} + k_{\nu} k_{\beta} \eta_{\mu \alpha} \right)]
\end{align}
and
\begin{align}
\frac{1}{\bar{\epsilon}} \equiv \frac{1}{\epsilon} - \gamma + \log \sqrt{4\pi}
\end{align}

with $2\epsilon = 4-d$.

In order to write the effective action, we transition back to position space. The momentum factors turn into derivatives acting on the external field. For example, the divergent term can be written as
\begin{align}
S_{div} = \frac{1}{3840\pi^2}\frac{1}{\bar{\epsilon}} \int d^4x \left(2 \partial_{\mu} \partial_{\nu} h^{\mu\nu} \partial_{\alpha} \partial_{\beta} h^{\alpha \beta} + \frac{3}{2} \partial^2 h \partial^2 h + \frac{1}{2} \partial^2 h_{\mu\nu} \partial^2 h^{\mu\nu} - 3 \partial_{\mu} \partial_{\nu} h^{\mu\nu} \partial^2 h - \partial_{\mu} \partial^{\nu} h_{\alpha\nu} \partial^{\mu} \partial_{\beta} h^{\beta\alpha} \right)\, .
\end{align}

\noindent The divergent contribution to the effective action goes into the renormalization of local operators in the gravitional action. Counting the number of derivatives in the above expression shows that the local operator we seek is composed of terms proportional to $R^2$. Hence, we seek the expansions of the different invariants up to second order in $h$.
\begin{align}
\nonumber
R &= -\partial_{\mu}\partial_{\nu} h^{\mu\nu} + \partial^2 h \\
R^2 &= \partial_{\mu} \partial_{\nu} h^{\mu\nu} \partial_{\alpha} \partial_{\beta} h^{\alpha\beta} - 2 \partial^2 h \partial_{\mu} \partial_{\nu} h^{\mu\nu} + \partial^2 h \partial^2 h
\end{align}
and
\begin{align}\label{rmunu}
\nonumber
R_{\mu\nu} &= \frac{1}{2} \left(-\partial_{\mu}\partial^{\alpha} h_{\alpha\nu} - \partial_{\nu}\partial^{\alpha} h_{\alpha\mu} + \partial^2 h_{\mu\nu} + \partial_{\mu}\partial_{\nu} h \right)\\
2 R_{\mu\nu} R^{\mu\nu} &= \frac{1}{2} \partial^2 h \partial^2 h + \frac{1}{2} \partial^2 h_{\mu\nu} \partial^2 h^{\mu\nu} + \partial_{\mu} \partial_{\nu} h^{\mu\nu} \partial_{\alpha} \partial_{\beta} h^{\alpha\beta} - \partial^2 h \partial_{\mu} \partial_{\nu} h^{\mu\nu} - \partial_{\mu} \partial_{\nu} h^{\mu\alpha} \partial^{\beta} \partial^{\nu} h_{\beta\alpha}\, .
\end{align}

\noindent Note that we have freely integrated by parts in these expressions. The gravitational effective Lagrangian is
\begin{align}
S = \int d^4x \sqrt{g} \left(\frac{1}{16 \pi G} R + c_1 R^2 + c_2 R_{\mu\nu} R^{\mu\nu}\right)\, .
\end{align}
Matching with the perturbative calculation allows us to identify the renormalized coupling constants as
\begin{align}
c_1 &= c^r_1(\mu) - \frac{1}{3840 \pi^2}\left(\frac{1}{\epsilon} - \gamma + \log \sqrt{4\pi} \right)\\
c_2 &= c^r_2(\mu) - \frac{1}{1920 \pi^2}\left(\frac{1}{\epsilon} - \gamma + \log \sqrt{4\pi} \right)\, .
\end{align}

Notice the explicit scale-dependence of the renormalized parameters which ensures the scale-independence of the effective action.\\

The non-local part of the effective action follows closely from the divergent part because the coefficient of $\log (-k^2)$ is uniquely tied to the
divergent $1/{\bar{\epsilon}}$ term. Following the logarithm in the transition to coordinate space, we find
\begin{align}\label{nlaction}
\nonumber
S_{non-local} =& \frac{1}{3840 \pi^2} \int d^4x  \int d^4y  [\partial^2 h(x) \bar{\mathfrak L}(x-y) \partial^2 h(y) + \partial_{\mu} \partial_{\nu} h^{\mu\nu}(x) \bar{\mathfrak L}(x-y) \partial_{\alpha} \partial_{\beta} h^{\alpha\beta}(y) - \partial^2 h(x) \bar{\mathfrak L}(x-y) \partial_{\mu} \partial_{\nu} h^{\mu\nu}(y)\\\nonumber
-& \partial_{\mu} \partial_{\nu} h^{\mu\nu}(x) \bar{\mathfrak L}(x-y) \partial^2 h(y) + \partial_{\mu} \partial^{\nu} h_{\nu\alpha}(x) \bar{\mathfrak L}(x-y) \partial^{\mu} \partial_{\beta} h^{\alpha\beta}(y) + \partial_{\mu} \partial^{\nu} h_{\nu\alpha}(x) \bar{\mathfrak L}(x-y) \partial^{\alpha} \partial_{\beta} h^{\mu\beta}(y)\\\nonumber
-&  \partial_{\mu} \partial^{\nu} h_{\nu\alpha}(x) \bar{\mathfrak L}(x-y) \partial^2 h^{\mu\alpha}(y) - \partial_{\mu} \partial^{\nu} h_{\nu\alpha}(x) \bar{\mathfrak L}(x-y) \partial^{\mu} \partial^{\alpha} h(y) - \partial^2 h_{\mu\nu}(x) \bar{\mathfrak L}(x-y) \partial^{\mu} \partial_{\beta} h^{\nu\beta}(y)\\\nonumber
-& \partial_{\mu} \partial_{\nu} h(x) \bar{\mathfrak L}(x-y) \partial^{\mu} \partial_{\beta} h^{\nu\beta}(y) + \frac{1}{2} \partial^2 h_{\mu\nu}(x) \bar{\mathfrak L}(x-y) \partial^2 h^{\mu\nu}(y) + \frac{1}{2} \partial^2 h_{\mu\nu}(x) \bar{\mathfrak L}(x-y) \partial^{\mu}\partial^{\nu}h(y)\\
+& \frac{1}{2} \partial_{\mu} \partial_{\nu} h(x) \bar{\mathfrak L}(x-y) \partial^2 h^{\mu\nu}(y) + \frac{1}{2} \partial_{\mu}\partial_{\nu}h(x) \bar{\mathfrak L}(x-y) \partial^{\mu}\partial^{\nu} h(y)]
\end{align}
where
\begin{align}\label{noncausal}
\bar{\mathfrak L}(x-y) = - \int \frac{d^4k}{(2\pi)^4} e^{-i k \cdot (x-y)} \log \left(\frac{-k^2}{\mu^2} \right)\, .
\end{align}
We note that each term in the momentum-space expression contributes to more than one term in the above position-space expression, so it needs some work to pass to Eqn. (\ref{nlaction}). Using the curvature expansions listed above, we easily realize a possible non-linear form of the non-local action
\begin{align}
S_{non-local} = \frac{1}{3840 \pi^2} \int d^4x  \int d^4y \, (\sqrt{g(x)}\sqrt{g(y)})^\frac12 \left[R(x) \bar{\mathfrak L}(x-y) R(y) + 2 R^{\mu}_{\nu}(x) \bar{\mathfrak L}(x-y) R^{\nu}_{\mu}(y) \right]\, .
\label{firsttime}
\end{align}
We note that the perturbative calculation alone does not enable us to differentiate between alternate forms of the non-linear completion which differ by application of the Gauss-Bonnet identity. We will remedy this issue and also discuss the nature of the approximation implied by this expression in Sec. 4. Note that the $\log \mu^2$ portion of $\bar{\mathfrak L}(x-y)$ corresponds to a delta function and hence is a finite local addition to $c_1$ and $c_2$. For $N$ scalar fields, the actions $S_{div}$ and $S_{nonl-local}$ are multiplied by a factor of $N$.

\section{Causal behavior}

The effective action of the previous section is {\em not} appropriate for generating causal effects in the equations of motion. The reason is
that the Feynman propagators involve both advanced and retarded solutions, and any variation of the effective action with respect to a field at time $t$
will involve the non-local effects both before and after $t$. This is appropriate for scattering amplitudes but not for the equations of motion. Rather
one needs to calculate the effects of the loops on the equations of motion using the  in-in (or Schwinger-Keldysh or closed-time-path) formalism \cite{keldysh, Barvinsky:1987uw, Vilkovisky:2007ny, Campos:1993ug, Jordan:1986ug, Calzetta:1986ey, Maciejko}, which
is designed to produce causal behavior. This is relatively more complicated and unfamiliar than usual perturbation theory. However, Bavinsky and Velkovisky \cite{Barvinsky:1987uw, Vilkovisky:2007ny} suggest the simple prescription  - that one merely varies the effective action (which they calculate in Euclidean space) and then afterwards imposes causal behavior or scattering behavior on the final result when one writes the answer in Lorentzian space. We perform the calculation below and confirm the validity of their prescription. The reader who is not interested in the details can skip to the results of Eqs. (\ref{expectation}), (\ref{causallog}) and (\ref{principal}), which are reasonably intuitive.

The in-in formalism deals not with the effective action but with expectation values. It is well known that the variation of the effective action yields the energy-momentum tensor of the quantum fields, and hence our strategy is to use the in-in formalism to calculate the {\em causal} energy-momentum tensor. The set-up of the formalism is laid out in the appendix, and our starting point is Eqn. (\ref{exval})
\begin{align}
\langle \mathcal{O}(t) \rangle &= ~_I\langle \Phi(-\infty) | S^{\dagger}(t,-\infty) \mathcal{O}_I(t) S(t,-\infty) |\Phi(-\infty) \rangle_I\, .
\end{align}
It is very useful to insert the identity operator in the form $S^{\dagger}(\infty,t)\, S(\infty,t) = 1$ to the left of the operator
\begin{align}
\langle \mathcal{O}(t) \rangle &= ~_I\langle \Phi(-\infty) | S^{\dagger}(\infty,-\infty) T\left[\mathcal{O}_I(t) S(\infty,-\infty)\right] |\Phi(-\infty) \rangle_I\, .
\end{align}
One then obtains various propagators - the normal Feynman propagators associated with purely time-ordered contrations, and others associated with mixed contractions as will be explicitly shown below.

For our case, we are calculating the expectation value of the energy momentum tensor to lowest order in the external field $h_{\mu\nu}$. Hence, we only have two bubble diagrams each with two propagators where one space-time point is the observation time. The first diagram arises from the $\mathcal{O}(h_{\mu\nu})$ term in $S(\infty,-\infty)$ and therefore contains the usual Feynman propagators. One obtains the non-local part of the expectation value
\begin{align}\label{emtfin}
\nonumber
\langle T^{NL}_{\mu\nu}(x) \rangle = \frac{1}{3840\pi^2} \int \frac{d^4k}{(2\pi)^4} &e^{-i k\cdot x}\, \log \left(\frac{-k^2}{\mu^2}\right)  h^{\alpha\beta}(-k) \,[8 k_{\mu}k_{\nu}k_{\alpha}k_{\beta} - k^2 ( 6 k_{\alpha}k_{\beta} \eta_{\mu\nu} + 6 k_{\mu}k_{\nu} \eta_{\alpha\beta} + k_{\nu}k_{\beta} \eta_{\mu\alpha} \\
&+ k_{\alpha}k_{\nu} \eta_{\mu\beta} + k_{\mu}k_{\beta} \eta_{\alpha\nu} + k_{\alpha}k_{\mu} \eta_{\beta\nu} )
+ k^4 ( \eta_{\mu\alpha}\eta_{\nu\beta} +\eta_{\mu\beta}\eta_{\alpha\nu} + 6 \eta_{\mu\nu}\eta_{\alpha\beta})]
\end{align}
where
\begin{align}
h^{\alpha\beta}(-k) = \int d^4y \, e^{ik\cdot y}h^{\alpha\beta}(y)\, .
\end{align}
This can be obtained either by direct calculation or by varying the effective action of the previous section. If we specialize to gravitational fields $h_{\mu\nu}(x)$ which are independent of spatial coordinates, we have
\begin{align}
\nonumber
\langle T^{NL}_{\mu\nu}(t) \rangle = \frac{1}{3840\pi^2} \, \int \frac{d\omega}{2\pi} & e^{- i \omega t}\, \left[ \log \left(\frac{-\omega^2}{\mu^2}\right) \right]\, h^{\alpha\beta}(-\omega) \,[8 k_{\mu}k_{\nu}k_{\alpha}k_{\beta} - k^2 (  6 k_{\alpha}k_{\beta} \eta_{\mu\nu} + 6 k_{\mu}k_{\nu} \eta_{\alpha\beta} \\
&+ k_{\nu}k_{\beta} \eta_{\mu\alpha} + k_{\alpha}k_{\nu} \eta_{\mu\beta} + k_{\mu}k_{\beta} \eta_{\alpha\nu} + k_{\alpha}k_{\mu} \eta_{\beta\nu} )
+ k^4 ( \eta_{\mu\alpha}\eta_{\nu\beta} +\eta_{\mu\beta}\eta_{\alpha\nu} + 6 \eta_{\mu\nu}\eta_{\alpha\beta})]
\end{align}
where now the momentum is purely temporal $k^{\mu} = (\omega,\vec{0})$ and
\begin{align}
h^{\alpha\beta}(-\omega) = \int dt^{\prime}\, e^{i\omega t^{\prime}}\, h^{\alpha\beta}(t^{\prime})\, .
\end{align}
Note that this result displays non-causal behavior because it is sensitive to times both before and after t.

The second diagram arises from the $\mathcal{O}(h_{\mu\nu})$ term in $S^{\dagger}(\infty,-\infty)$. To calculate such diagram, the algebra of contractions needs a modification to Wick's theorem to incorporate anti-time-ordered product of operators. The details of the construction is included in the appendix. Only the last two terms in Eqn. (\ref{desired}) involving products of positive-frequency Wightman functions contribute to the calculation. We denote this particular Wightman function by an underline
\begin{align}
\underline{\phi(x) \phi(y)} \equiv [\phi^+(x),\phi^-(y)] &= \langle 0| [\phi^+(x),\phi^-(y)] |0 \rangle = \langle 0|\phi^+(x) \phi^-(y)|0 \rangle = \langle 0|\phi (x) \phi(y)|0 \rangle
\end{align}
and it explicitly reads
\begin{align}
\underline{\phi(x) \phi(y)} = 2\pi \, \int \frac{d^4p}{(2\pi)^4}\, \theta(p^0)\, \delta(p^2) \, e^{-i p \cdot (x - y)}\, .
\end{align}
The result is a simple addition to the expectation value, with a total result that reads
\begin{align}
\nonumber
\langle T^{NL}_{\mu\nu}(t) \rangle = \frac{1}{3840\pi^2} \, \int \frac{d\omega}{2\pi} \, & e^{- i \omega t}\, \left[ \log \left(\frac{-\omega^2}{\mu^2}\right) + 2i \pi \theta(-\omega) \right]\, h^{\alpha\beta}(-\omega) \,[8 k_{\mu}k_{\nu}k_{\alpha}k_{\beta} - k^2 (  6 k_{\alpha}k_{\beta} \eta_{\mu\nu} + 6 k_{\mu}k_{\nu} \eta_{\alpha\beta} \\
&+ k_{\nu}k_{\beta} \eta_{\mu\alpha} + k_{\alpha}k_{\nu} \eta_{\mu\beta} + k_{\mu}k_{\beta} \eta_{\alpha\nu} + k_{\alpha}k_{\mu} \eta_{\beta\nu} )
+ k^4 ( \eta_{\mu\alpha}\eta_{\nu\beta} +\eta_{\mu\beta}\eta_{\alpha\nu} + 6 \eta_{\mu\nu}\eta_{\alpha\beta})]\, .
\end{align}

Again we transform the above expression to real space, with momentum factors turning into derivatives. This yields
\begin{align}
\langle T^{NL}_{\mu\nu}(t) \rangle = \int dt^{\prime}\, \mathfrak L(t-t^{\prime})\, \mathcal{D}_{\mu\nu\alpha\beta} h^{\alpha\beta}(t^{\prime})
\label{expectation}
\end{align}
where
\begin{align}
\mathcal{D}_{\mu\nu\alpha\beta} =\frac{1}{3840\pi^2} \, &[8 \partial_{\mu}\partial_{\nu}\partial_{\alpha}\partial_{\beta} + \partial^4 ( \eta_{\mu\alpha}\eta_{\nu\beta} + \eta_{\mu\beta}\eta_{\alpha\nu} + 6 \eta_{\mu\nu}\eta_{\alpha\beta})  \nonumber \\
&- \partial^2 (  6 \partial_{\alpha}\partial_{\beta} \eta_{\mu\nu} + 6 \partial_{\mu}\partial_{\nu} \eta_{\alpha\beta} + \partial_{\nu}\partial_{\beta} \eta_{\mu\alpha} + \partial_{\alpha}\partial_{\nu} \eta_{\mu\beta} + \partial_{\mu}\partial_{\beta} \eta_{\alpha\nu} + \partial_{\alpha}\partial_{\mu} \eta_{\beta\nu})]
\end{align}
and where we have identified our key non-local function
\begin{align}
\mathfrak L(t-t^{\prime}) = \int_{-\infty}^{\infty} \frac{d\omega}{2\pi} \, e^{-i\omega(t - t^{\prime})} \, \left[ \log \left(\frac{-\omega^2}{\mu^2}\right) + 2i \pi \theta(-\omega) \right]\, .
\label{causallog}
\end{align}

In order to evaluate this integral, we first note
\begin{align}
\log \left(\frac{-\omega^2}{\mu^2}\right) = \log \left(\frac{\omega^2}{\mu^2}\right) - i\pi ,\quad -i\pi + 2i\pi \theta(-\omega) = -i\pi \, \text{sgn}(\omega)
\end{align}
hence
\begin{align}\label{l}
\nonumber
\mathfrak L(t-t^{\prime}) &= - \, 2 \int_{-\infty}^{\infty} \frac{d\omega}{2\pi} \, e^{-i\omega(t - t^{\prime})} \, \left[ \log \left(\frac{\mu}{|\omega|}\right) + \frac{i \pi}{2} \text{sgn}(\omega) \right]\\
&= -2\, \mathcal{P} \, \frac{\theta(t-t^{\prime})}{t-t^{\prime}}\, .
\end{align}
Here $\mathcal{P}$ denotes the principal value distribution \cite{Frob:2012ui} defined by
\begin{align}
\mathcal{P} \, \frac{\theta(t-t^{\prime})}{t-t^{\prime}} = \lim_{\epsilon \rightarrow 0} \, \left[\frac{\theta(t-t^{\prime}-\epsilon)}{t-t^{\prime}} + \delta(t-t^{\prime})\left(\log (\mu \epsilon) + \gamma \right) \right]\, .
\label{principal}
\end{align}
Unlike Eqn. (\ref{noncausal}), this function is clearly causal and real. It also provides a precise definition of how the non-local integration is to be performed as the term with the delta function yields the desired feature that the non-local effect is finite.
This result verifies the Bavinsky-Velkovisky procedure of varying the effective action and then simply imposing causal behavior.

\section{Non-linear completion and quasi-local form}
The perturbative analysis gives us a reference for the form of the non-local quantum effects and the precise causal prescription. In order to have a more complete description appropriate for application to FLRW cosmology, we can match to the work by Barvinsky, Vilkovisky and collaborators \cite{Barvinsky:1985an, Barvinsky:1995jv, Barvinsky:1994ic, Avramidi:1990je, gospel}. These authors have explored non-local aspects of the heat kernel expansion and expressed the results in quasi-local form. Normally the heat kernel methodology is used to capture local quantum effects. For example, the second coefficient in the expansion of the one-loop effective action, commonly called $a_2(x)$, gives the divergent terms that go into the renormalization of the effective Lagrangian quadratic in curvature invariants. For massless fields, this is the only one-loop divergence. However, the asymptotic form of the heat kernel expansion also reveals non-analytic terms. These authors study the non-analytic terms in Euclidean space and display the results using quasi-local actions of the form

\begin{align}
S_{QL} = \int d^4x \, \sqrt{g} \, \left[R \log\left(\frac{\Box}{\mu^2}\right) R \right]\, .
\end{align}

\noindent Here, $\Box$ is the d'Alembertian operator as before. Despite the fact that this appears to be expressed in local form, we show below that matching to the perturbative calculation of the preceding sections confirms that it corresponds to a non-local effect. The quasi-local forms provide a non-linear covariant completion of the perturbative calculation.\\

If we resolve the operator $\log\left(\frac{\Box}{\mu^2}\right)$ by introducing position space eigenstates we find
\begin{align}
S_{QL} = \int \, d^4x \, \sqrt{g(x)} R(x) \, \int \, d^4y \sqrt{g(y)} \, \langle x | \log\left(\frac{\Box}{\mu^2}\right) | y \rangle R(y)\, .
\end{align}

\noindent Here the states are normalized covariantly
\begin{align}
\langle x | y \rangle = \frac{\delta^{(4)}(x-y)}{\left(\sqrt{g(y)}\sqrt{g(x)}\right)^{1/2}}\, .
\end{align}

\noindent If we also define

\begin{align}
\langle x | \log\left(\frac{\Box}{\mu^2}\right) | y \rangle = \left(\sqrt{g(y)}\sqrt{g(x)}\right)^{-1/2} L(x,y;\mu)
\end{align}

\noindent we can write the action in explicitly non-local form
\begin{align}
S_{NL} = \int \, d^4x \, \int d^4y \, \sqrt{g(x)}^{1/2} \, R(x) \, L(x,y;\mu) \, \sqrt{g(y)}^{1/2} \, R(y)\, .
\end{align}

\noindent Again, we note that the $\log \mu$ dependence in these equations corresponds to a local effect. Here, we see that replacing the covariant d'Alembertian in Eqn. (44) by its Minkowski couterpart yields the first term in Eqn. (\ref{firsttime}). \\

There are three terms in the general non-local Lagrangian. Reverting temporarily to quasi-local form, these can be written as
\begin{align}\label{generalL}
S_{QL} = \int d^4x \sqrt{g} \left(\alpha R \log\left(\frac{\Box}{\mu_{\alpha}^2}\right) R + \beta R_{\mu\nu} \log\left(\frac{\Box}{\mu_{\beta}^2}\right) R^{\mu\nu} + \gamma R_{\mu\nu\alpha\beta} \log\left(\frac{\Box}{\mu_{\gamma}^2}\right) R^{\mu\nu\alpha\beta} \right)
\end{align}
where $\alpha, \beta, \gamma$ are numerical coefficients which we will display below. We allow for the possibility that the renormalization scales are different for the three terms as the coupling constants of the local Lagrangian could be measured at different scales. For local terms, there are only two quadratic invariants to be considered due to the Gauss-Bonnet identity which holds strictly in four dimensions
\begin{align}
\int d^4x \, \sqrt{g} \, R_{\mu\nu\alpha\beta} R^{\mu\nu\alpha\beta} = \int d^4x \, \sqrt{g} \, [4R_{\mu\nu}R^{\mu\nu} - R^2] + \text{total derivative}\, .
\end{align}

\noindent While Eqn. (\ref{generalL}) is simple and easy to apply, an alternate form reveals some interesting physics. For this form we employ the Weyl tensor in four dimensions
\begin{align}
C_{\mu\nu\alpha\beta} = R_{\mu\nu\alpha\beta} - \frac{1}{2} \left(g_{\mu\alpha} R_{\nu\beta} + g_{\mu\beta} R_{\nu\alpha}  + g_{\nu\alpha} R_{\mu\beta} - g_{\nu\beta} R_{\mu\alpha}\right) + \frac{1}{6} R \left(g_{\mu\alpha} g_{\nu\beta} - g_{\mu\beta}g_{\nu\alpha}\right)
\end{align}
to rewrite
\begin{align}
\nonumber
S_{QL} = \int d^4x \, \sqrt{g} \, & \bigl[\bar{\alpha} R \log\left(\frac{\Box}{\mu_1^2}\right) R + \bar{\beta} C_{\mu\nu\alpha\beta} \log\left(\frac{\Box}{\mu_2^2}\right) C^{\mu\nu\alpha\beta} + \bar{\gamma} \bigl(R_{\mu\nu\alpha\beta}\log\left({\Box}\right)R^{\mu\nu\alpha\beta} - 4 R_{\mu\nu}\log\left({\Box}\right) R^{\mu\nu}\\
&+ R \log\left({\Box}\right) R \bigr)\bigr]\, .
\label{GB}
\end{align}
This form has several theoretical advantages. Here the last term, similar in structure to the Gauss-Bonnet term, does not have any $\mu$ dependence because its local form does not contribute to the equations of motion. The FLRW metric that we use below is conformally flat and thus its Weyl tensor vanishes. Thus the second term will not contribute to our cosmological application. In turn this tells us that the cosmology study dependence on local short distance physics comes through the first term only, and there is only one parameter $\mu_1 \equiv \mu$ which describes this local term. In addition this first term is not generated by conformally invariant field theories (fermions, photons and conformally coupled scalars) and their quantum effects will be purely non-local. The coefficients in these two different bases are related by
\begin{align}
\alpha = \bar{\alpha} +\frac{\bar{\beta}}{3}+ \bar{\gamma},\quad \beta = -2\bar{\beta}-4 \bar{\gamma}, \quad \gamma = \bar{\beta} + \bar{\gamma}\, .
\end{align}

We can identify the coefficients in the non-local Lagrangian because the logarithms are tied to the divergences in the one-loop effective action, as shown by the perturbative calculation. The latter have been calculated in the background field method, and results are known before the Gauss-Bonnet identity has been applied\footnote{This background field method resolves the problem of identifying the complete form of the non-linear completion that we had in discussing
Eq. (\ref{firsttime}).}. For example, the divergent effective Lagrangian for a massless field reads

\begin{align}
\mathcal{L}_{div} = \sqrt{|g|}\, \frac{a_2(x)}{16\pi^2 \, \epsilon}\, .
\end{align}

\noindent The coefficient $a_2(x)$ is known for scalars, fermions and photons \cite{Birrell:1982ix, deser}
\begin{align}
a^S_2(x) &= \frac{1}{180} \left( \frac{5}{2}R^2 - R_{\mu\nu}R^{\mu\nu} + R_{\mu\nu\alpha\beta}R^{\mu\nu\alpha\beta} \right)\\
a^F_2(x) &= \frac{1}{360}\left(- 5 R^2 + 8 R_{\mu\nu}R^{\mu\nu} + 7 R_{\mu\nu\alpha\beta}R^{\mu\nu\alpha\beta}\right)\\
a^V_2(x) &= \frac{-1}{180} \left(20 R^2 - 86 R_{\mu\nu}R^{\mu\nu} + 11 R_{\mu\nu\alpha\beta}R^{\mu\nu\alpha\beta}\right)\, .
\end{align}
Here, the result for fermions assumes a four-component spinor field. The result for the massless vector field also includes the ghost contribution, which is twice the scalar field result with an appropriate minus sign. Finally, the classic paper by 't Hooft and Veltman \cite{'tHooft:1974bx} gave the result for gravitons only after using the Gauss-Bonnet relation, but the general result has since been calculated, see e.g. \cite{Buchbinder:2012wb}. This enables us to read off the result for gravitons which also includes the ghost contribution

\begin{align}
a^G_2(x) = \frac{215}{180} R^2 - \frac{361}{90} R_{\mu\nu}R^{\mu\nu} + \frac{53}{45} R_{\mu\nu\alpha\beta}R^{\mu\nu\alpha\beta}\, .
\end{align}

\noindent In table (\ref{coeff1}), we collect the coefficients of different fields.

\begin{table}
\begin{tabular}{| c | c | c | c | c | c | c |}
\hline
 & $\alpha$ & $\beta$ & $\gamma$ & $\bar{\alpha}$ & $\bar{\beta}$ & $\bar{\gamma}$\\
 \hline
 \text{Scalar} & $ 5(6\xi-1)^2$ & $-2 $ & $2$   & $ 5(6\xi-1)^2$ & $3 $ & $-1$   \\
 \hline
 \text{Fermion} & $-5$ & $8$ & $7 $ & $0$ & $18$ & $-11 $\\
 \hline
 \text{Vector} & $-50$ & $176$ & $-26$ & $0$ & $36$ & $-62$\\
 \hline
 \text{Graviton} & $430$ & $-1444$ & $424$& $90$ & $126$ & $298$\\
 \hline
\end{tabular}
\caption{Coefficients of different fields. All numbers should be divided by $11520\pi^2$.}
\label{coeff1}
\end{table}

The results are shown for a scalar with a coupling $\xi R\phi^2$ and the parameter $\xi$ enters the $\alpha$ couplings
\begin{align}
\alpha =     \bar{\alpha} = \frac{(6\xi -1)^2}{2304\pi^2}
\end{align}
with $\beta,~\gamma,~\bar{\beta},~\bar{\gamma}$ independent of $\xi$. Unless stated otherwise, our results are presented for a minimally coupled scalar ($\xi=0$), while a conformally coupled scalar has $\xi=1/6$. For conformally invariant fields the coefficient $\bar{\alpha}$ will vanish. Because the FLRW metric is conformally flat, the coupling $\bar{\beta}$ does not contribute to our analysis as mentioned previously. This leaves only the coefficient $\bar{\gamma}$ as the active parameter. For $N_S$ scalars, $N_f$ fermions and $N_V$ gauge bosons, this coupling has the value
\begin{align}
\bar{\gamma} = -\frac{1}{11520 \pi^2}[N_S+11N_f +62N_V]\, .
\end{align}
Note that all conformally invariant matter fields carry the same sign of $\bar{\gamma}$ and will have similar effects, differing just in magnitude. Moreover, this case is independent of the parameter $\mu$ because the Gauss-Bonnet non-local term (the one proportional to $\bar{\gamma}$) has no local contribution to the equations of motion.

Finally, we can also add up the contributions of all the SM particles (plus the graviton) to find effective SM coefficients which are calculated as follows
\begin{align}
\alpha_{SM} = N_S \alpha_S + N_l \alpha_F + N_c N_q \alpha_F + N_V \alpha_V + \alpha_G
\end{align}
and likewise for $\beta$ and $\gamma$. Here, we have broken the fermion contribution up into quark and lepton terms $N_f= N_l+N_c N_q$ where $N_l$ is the number of leptons, $N_q$ and $N_c$ are the numbers of quarks and colors respectively. For the standard model with a minimally coupled Higgs, these numbers read
\begin{align}
N_S = 4, \quad N_l = 6, \quad N_c = 3 , \quad N_q = 6, \quad N_V = 12\, .
\end{align}

\noindent Hence, for this case we find
\begin{align}
\alpha_{SM} = \frac{-7}{1152 \pi^2}, \quad \beta_{SM} = \frac{287}{1440 \pi^2} , \quad \gamma_{SM} = \frac{-17}{1440 \pi^2}
\end{align}
for the Standard Model particles alone, or also including gravitons
\begin{align}
\alpha_{SMG} = \frac{-3}{128 \pi^2}, \quad \beta_{SMG} = \frac{71}{960 \pi^2} , \quad \gamma_{SMG} = \frac{1}{40 \pi^2}\, .
\end{align}
For a conformally coupled Higgs field we find the conformally invariant result (without gravitons) $\bar{\alpha}_c=0$ and
\begin{align}
\bar{\gamma}_c = -\frac{253}{2880 \pi^2}\, .
\end{align}

\noindent Of course, we recognize that we expect to find new particles between the weak scale and the Planck scale, and so these numbers would likely be modified when the formalism is applied near the Planck scale.

\section{Non-local FLRW equations}

The equations of motion can be obtained by varying the effective action, specializing to the FLRW metric and then imposing causal prescription. We do that in this section, displaying the corresponding non-local effects in the FLRW equations.

However, this procedure cannot be done exactly because we do not know the full dependence of the non-local function $L(x,y;\mu)$ on the background metric. We could employ a procedure like the Riemann normal coordinates expasnion \cite{Birrell:1982ix}, in which propagators are expanded in powers of the curvature. A calculation of the loop diagram could then be used to provide an expansion of $L(x,y;\mu)$ involving further powers of the curvature through triangle and box diagrams. However, since the quasi-local action is already quadratic in the curvature, we will proceed by dropping such higher curvature terms and employing the approximation
\begin{align}
L(x,y;\mu) \approx \bar{{\mathfrak L}} (x-y)
\end{align}
when we pass to the non-local form of the action. This approximation confines our study to quadratic corrections to the gravitational action. Because the non-local function $\bar{{\mathfrak L}}(x-y)$ falls as $1/(t-t^{\prime})$ our approximation captures the behavior where the integrand is the largest, but will differ past the Hubble time where the integrated curvature becomes large.
With this approximation, the non-local function depends only on $|x-y|$ so that
\begin{align}
\frac{\partial}{\partial x} \bar{{\mathfrak L}}(x-y) = -\frac{\partial}{\partial y} \bar{{\mathfrak L}}(x-y)
\end{align}
allowing derivatives acting on $\bar{{\mathfrak L}}$ to be transferred to derivatives acting on the scale factor $a(t^{\prime})$.

The non-linear FLRW equations can be derived in one of two ways. One can vary $g_{\mu\nu}$ in general and then specialize to the FLRW metric. Equivalently one may use the general metric $ds^2 = f^2(t) dt^2 - a^2(t)d^2\bold{x}$, varying with respect to both $f$ and $a$ and then setting $f=1$ at the end. Either way we obtain the $0-0$ component of the modified equations of motion
\begin{align}
\frac{3 a \dot{a}^2 }{8\pi}+ N \left[ 6 (\sqrt{a} \, \ddot{a})_t \int \, dt^{\prime} \, \mathfrak L(t-t^{\prime}) \mathcal{R}_1 + 6 \left(\frac{\dot{a}^2}{\sqrt{a}}\right)_t \, \int \, dt^{\prime} \, \mathfrak L(t-t^{\prime}) \mathcal{R}_2 + 12 (\sqrt{a} \dot{a})_t \int \, dt^{\prime} \, \mathfrak L(t-t^{\prime}) \frac{d\, \mathcal{R}_3}{dt^{\prime}}\right] = a^3 \rho\, .
\end{align}

\noindent Here, $N$ represents the number of particles and the different functions read
\begin{align}
\mathcal{R}_1 &= - \sqrt{a} \ddot{a}(6 \alpha + 2 \beta + 2 \gamma) - \frac{\dot{a}^2}{\sqrt{a}} (6 \alpha + \beta)\\
\mathcal{R}_2 &= - \sqrt{a} \ddot{a} ( 12 \alpha + \beta - 2 \gamma) - \frac{\dot{a}^2}{\sqrt{a}} (12 \alpha  + 5 \beta + 6 \gamma)\\
\mathcal{R}_3 &= \sqrt{a}\ddot{a}(6 \alpha + 2 \beta + 2 \gamma) + \frac{\dot{a}^2}{\sqrt{a}}(6 \alpha + \beta)\, .
\end{align}
For mixed combinations of particles, $N$ can be absorbed in the definitions of $\alpha_{tot}, ~\beta_{tot}, ~\gamma_{tot}$ as described in the previous section. As described in Sec. 3, the equations of motion must use the causal non-local function
\begin{align}
\mathfrak L(t-t^{\prime}) = \lim_{\epsilon \rightarrow 0} \left[ \frac{\theta(t - t^{\prime}- \epsilon)}{t-t^{\prime}} + \delta(t-t^{\prime}) \log(\mu_R \, \epsilon) \right]
\end{align}
\noindent obtained therein and we absorbed Euler's constant into the renormalization scale $\mu_R$. We finally remind that in a covariant theory the space-space equation of motion is {\em not} an independent equation. This is not true in our case since we employed an approximation for the function $L(x,y;\mu)$ that manifestly breaks general covariance.

\section{Emergence of classical behavior}

In assessing the effects of the non-local behavior, we treat the new terms as a perturbation in the equation of motion. They have certainly been calculated as perturbations to the leading behavior, so this is a conservative approach. We will address the limits of such perturbative treatment in the final section.

In an expanding universe, the quantum effects are expected to be felt most in the early phases of expansion when the curvature is largest. In principle, these effects could change the character of the expansion, perhaps by an instability. In addition, the memory effect which is sensitive to past values of the curvature with the weight $1/(t-t^{\prime})$ could have an effect which builds up with time. Within our approximation, neither of these happens. We will explore the situation by 'switching on' the non-local effect at the Planck time. The evolution of the scale factor is influenced by the non-local effect very close to the Planck time. However, subsequent evolution turns essentially classical and the effect of non-local terms fades away.

We will treat both a dust-filled universe and a radiation-filled universe. We set $G=1$ in the numerical evaluation. The lower limit of the integrals is then taken to be $t_0 = 1$ which corresponds to the Planck time as mentioned earlier. In treating the new terms as a perturbation, we use the known classical solutions as input to the integrands, integrating up to the observation time $t$. This allows the integrals over time to be done by hand and converts the integro-differential equation into a simpler differential equation, albeit one with a reference back to a starting time $t_0$.

For a scalar field, we use the coefficients listed in the previous section to find the functions
\begin{align}
\mathcal{R}_1 = \frac{-1}{\pi^2} \left(\frac{\sqrt{a}\ddot{a}}{384} + \frac{7 \dot{a}^2}{2880 \sqrt{a}} \right), \quad \mathcal{R}_2 = \frac{-1}{\pi^2}\left(\frac{3\sqrt{a}\ddot{a}}{640} + \frac{31 \dot{a}^2}{5760 \sqrt{a}} \right), \quad \mathcal{R}_3 = \frac{1}{\pi^2} \left(\frac{\sqrt{a}\ddot{a}}{384} + \frac{7 \dot{a}^2}{2880 \sqrt{a}} \right)\, .
\end{align}

\noindent If we treat the matter input as dust, the classical solution is $a(t) = (t/t_0)^{2/3}$ and thus the $0-0$ equation of motion reads
\begin{align}
a \dot{a}^2 - \frac{N_S}{2430 \pi} \left(\frac{19 E_1(t;t_0)}{t^2_0 t} + \frac{26 E_2(t;t_0)}{t^2_0} \right) = \frac{8 \pi \rho_0}{3}\, .
\end{align}
We note that the normalization time is chosen to coincide with the initial time $t_0$, and hence the energy density is $\rho_0 = 1/(6 \pi t^2_0)$. We also defined the functions
\begin{align}
E_1(t;t_0) = \frac{\log(\mu_R t) + \log(t/t_0 - 1)}{t}, \quad E_2(t;t_0) = \frac{\log (\mu_R t) + \log(t/t_0 - 1) + (t/t_0-1)}{t^2}\, .
\end{align}

Results are shown in Figs. \ref{E100}-\ref{E104} for different numbers of scalar fields. In each case, the quantum correction provides an initial deviation from the straight classical behavior. However, as the scale factor evolves, the curvature decreases and the evolution is driven by the lowest order FLRW equation with the usual classical form. This is perhaps expected but indicates, at least within our approximations, that the quantum terms do not destabilize the evolution of the scale factor. One can see that increasing the number of scalars increases the magnitude of the quantum effect, but does not change the character of the effect. For these plots we have used $\mu_R=1$, but a reasonable range of other values of $\mu_R$ leads to qualitatively similar results.

We also show the case of pure graviton loops in Fig. \ref{Egrav}. This is qualitatively similar to that of scalars, with the graviton making a somewhat larger effect than would an individual scalar.

\begin{figure*}
\centering
\begin{tabular}{cc}
\epsfig {file=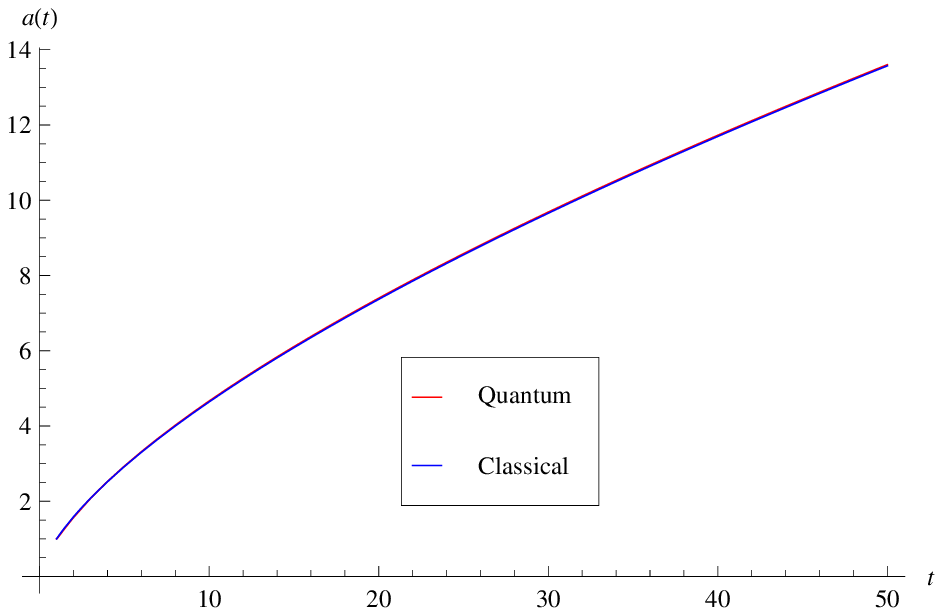,width=0.5\linewidth,clip=}&
\epsfig {file=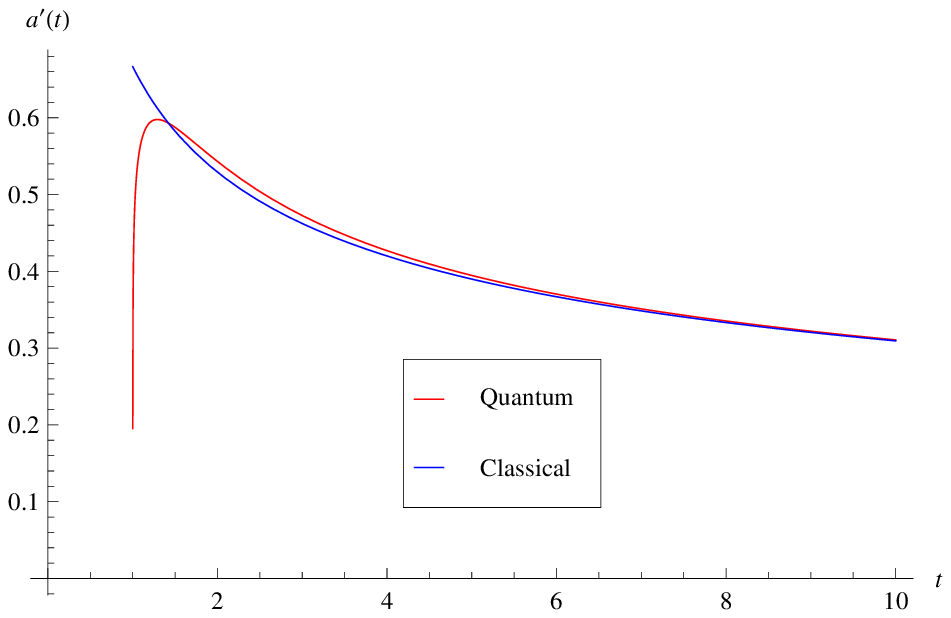,width=0.5\linewidth,clip=}
\end{tabular}
\caption{The evolution of the scale factor and its time derivative in an expanding dust filled universe for N=10.}
\label{E100}
\end{figure*}

\begin{figure*}
\centering
\begin{tabular}{cc}
\epsfig {file=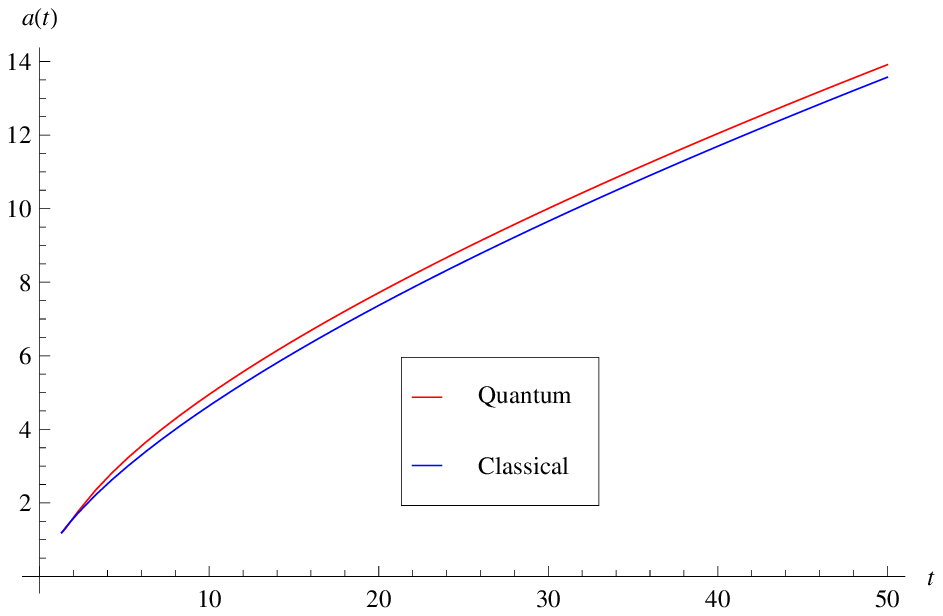,width=0.5\linewidth,clip=}&
\epsfig {file=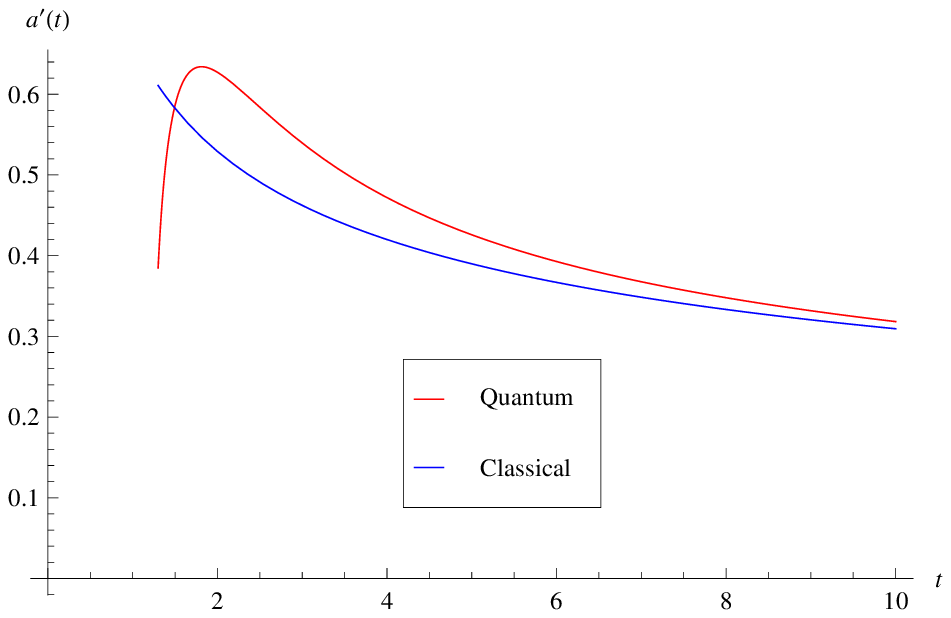,width=0.5\linewidth,clip=}
\end{tabular}
\caption{The evolution of the scale factor and its time derivative in an expanding dust-filled universe for N=100.}
\label{E10}
\end{figure*}

\begin{figure*}
\centering
\begin{tabular}{cc}
\epsfig {file=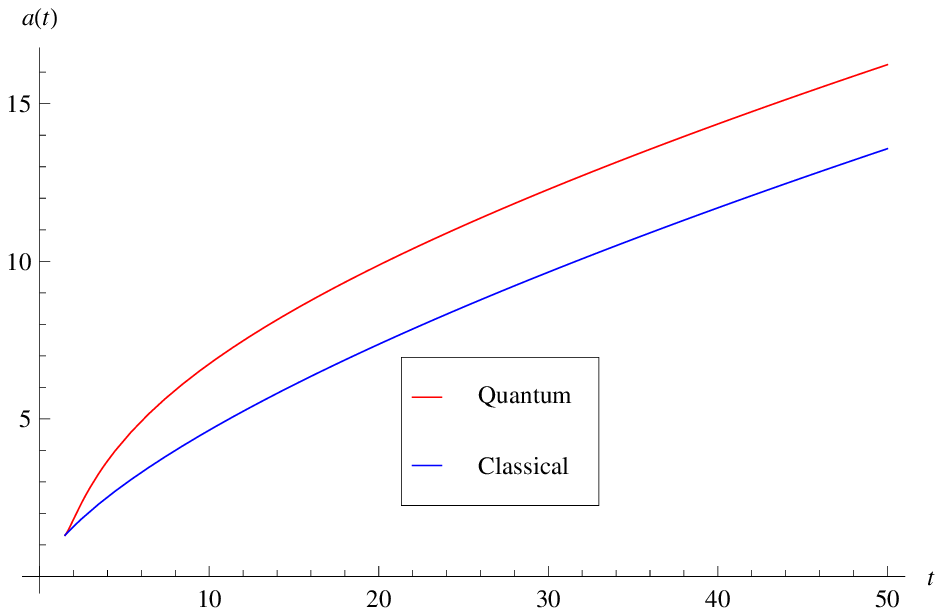,width=0.5\linewidth,clip=}&
\epsfig {file=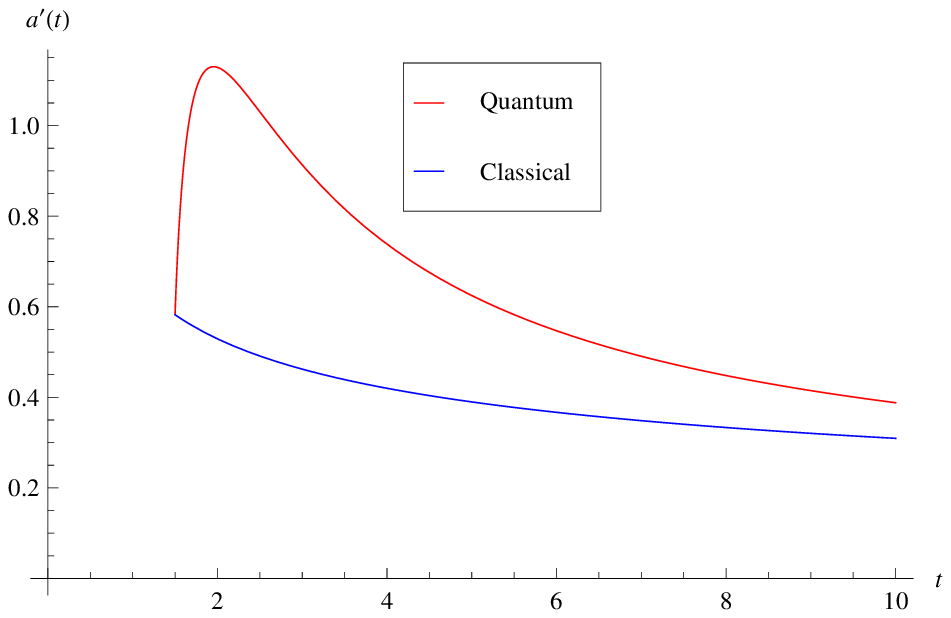,width=0.5\linewidth,clip=}
\end{tabular}
\caption{The evolution of the scale factor and its time derivative in an expanding dust-filled universe with $N=10^4$ scalar fields.}
\label{E104}
\end{figure*}

\begin{figure*}
\centering
\begin{tabular}{cc}
\epsfig {file=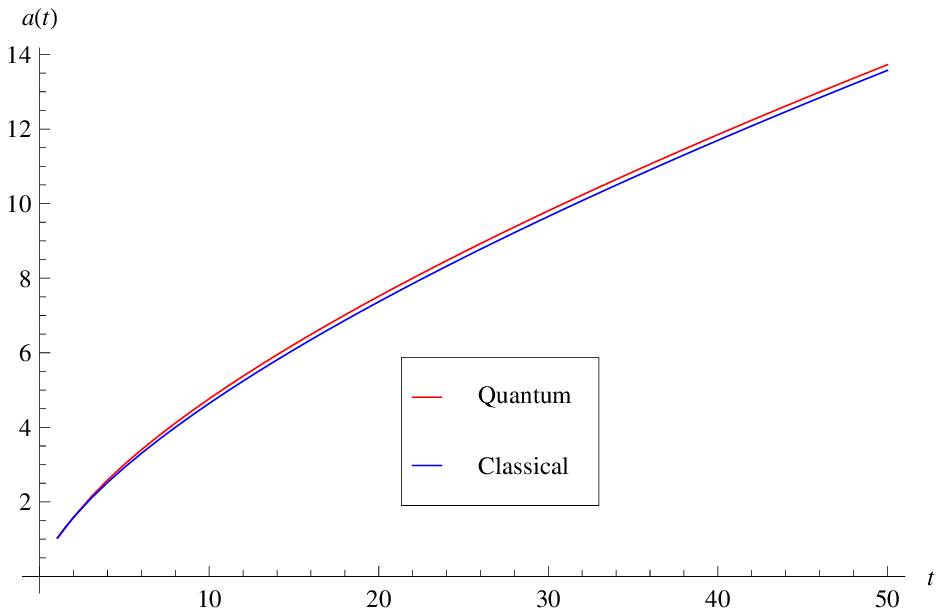,width=0.5\linewidth,clip=}&
\epsfig {file=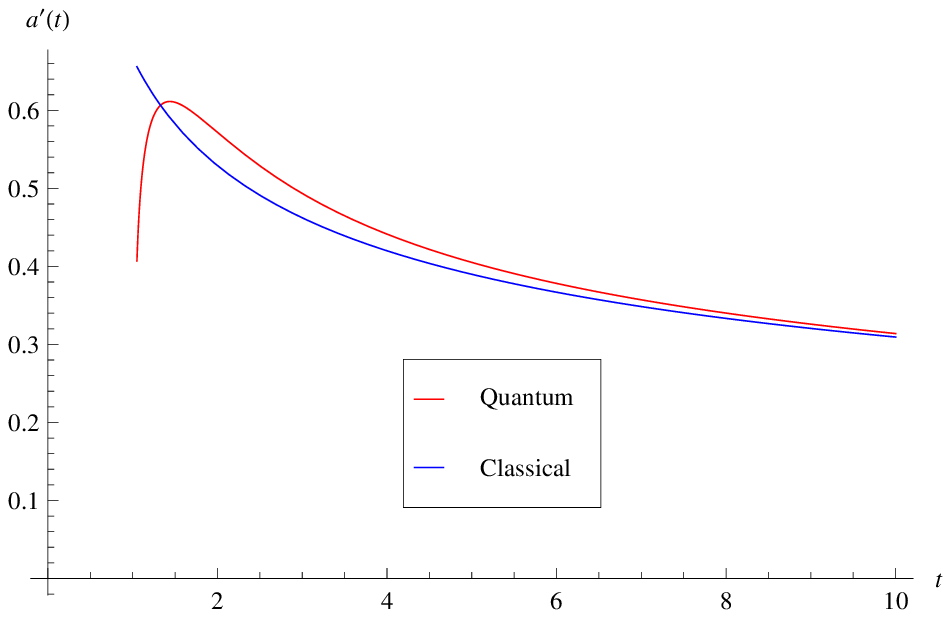,width=0.5\linewidth,clip=}
\end{tabular}
\caption{The evolution of the scale factor and its time derivative in an expanding dust-filled universe with quantum graviton loops.}
\label{Egrav}
\end{figure*}

\begin{figure*}
\centering
\begin{tabular}{cc}
\epsfig {file=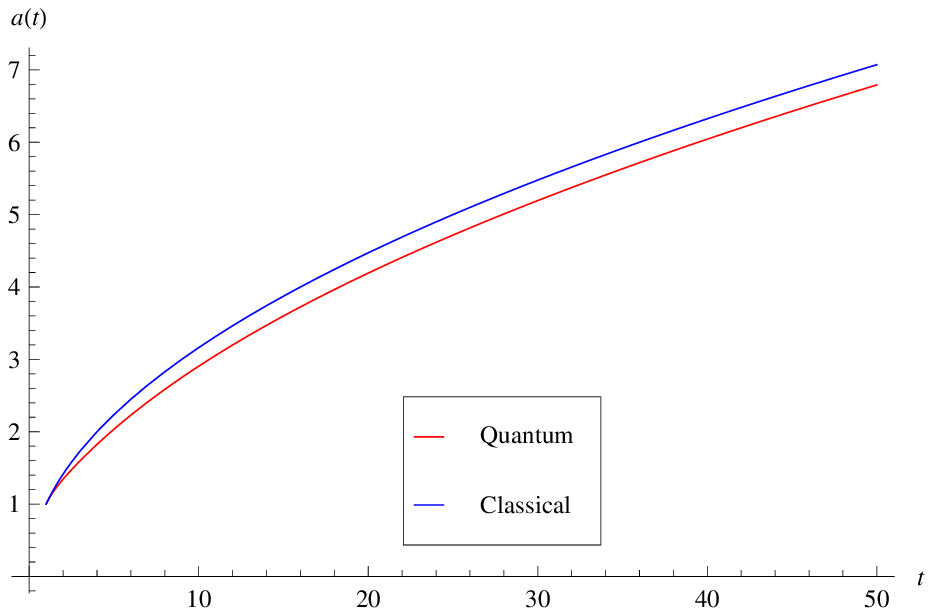,width=0.5\linewidth,clip=}&
\epsfig {file=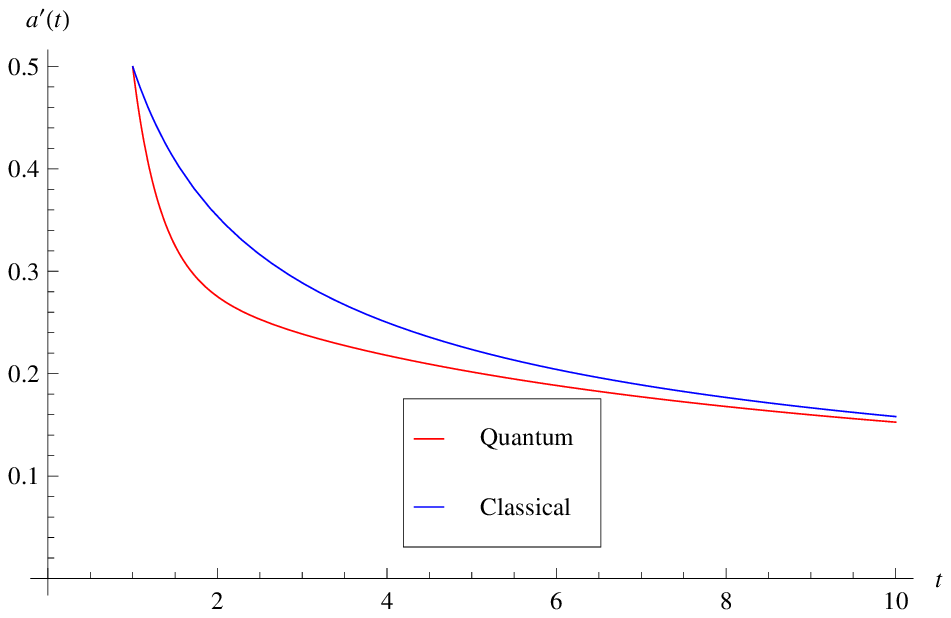,width=0.5\linewidth,clip=}
\end{tabular}
\caption{The evolution of the scale factor and its time derivative in an expanding radiation-filled universe with $N=10^3$ scalar fields.}
\label{Erad}
\end{figure*}

For a radiation dominated universe the situation is also interesting,
\begin{align}
a \dot{a}^2 - \frac{N_S}{1152 \pi } \left(\frac{E_{5/4}(t;t_0)}{t^{3/2}_0 t^{5/4}} - \frac{E_{9/4}(t;t_0)}{t^{3/2}_0 t^{1/4}}\right) = \frac{8 \pi \rho_0}{3a}\, .
\end{align}

\noindent In this case the energy density is $\rho_0 = 3/(32\pi t^2_0)$. The expansion functions read
\begin{align}
E_{5/4}(t,t_0) &= \frac{1}{t^{5/4}} \left[ \log(\mu_R t) + \log \left(\frac{t^{1/4} - t_0^{1/4}}{t^{1/4} + t_0^{1/4}} \right) + 4 \left(\frac{t}{t_0}\right)^{1/4} + 2 \arctan \left(\frac{t_0}{t}\right)^{1/4} + \log(8) - 4 - \frac{\pi}{2} \right]\\
E_{9/4}(t,t_0) &= \frac{1}{t^{9/4}} \left[ \log(\mu_R t) + \log \left(\frac{t^{1/4} - t_0^{1/4}}{t^{1/4} + t_0^{1/4}} \right) + 4 \left(\frac{t}{t_0}\right)^{1/4} + \frac{5}{4} \left(\frac{t}{t_0}\right)^{5/4} + 2 \arctan \left(\frac{t_0}{t}\right)^{1/4} + \log(8) - \frac{21}{4} - \frac{\pi}{2} \right]\, .
\end{align}
The equation of motion shows the interesting feature that the dependence on $\log\mu_R$ cancels out, which means that the effect is purely non-local. The reason is that the classical solution in the case of radiation $a(t)=(t/t_0)^{1/2}$ furnishes an exact solution to {\em local} quadratic gravity.
We show results for the expanding radiation universe with a thousand scalar fields in Fig. \ref{Erad}. The quantum effects are somewhat smaller in the radiation case, but have the same qualitative behavior as the dust-filled universe. Situations involving fermions, photons and gravitons are also quite similar and we do not display figures for each case.

Overall these results are satisfying in that the quantum corrections are well behaved and turn off as we enter the period of classical evolution.

\section{Contracting universe and the possibility of a bounce}
Of perhaps greater interest is the physics of a collapsing phase. Here the initial conditions are purely classical and the natural evolution bring the universe into the quantum regime. The classical evolution is headed towards a singularity - the big crunch. We will explore this case and see that within our approximation the quantum effects can lead to an avoidance of the singularity.
%%%%%%%%%%%%%%%%%%%%%%%%%
\begin{figure}[ht]
\centerline{
\includegraphics[width=0.7\textwidth]{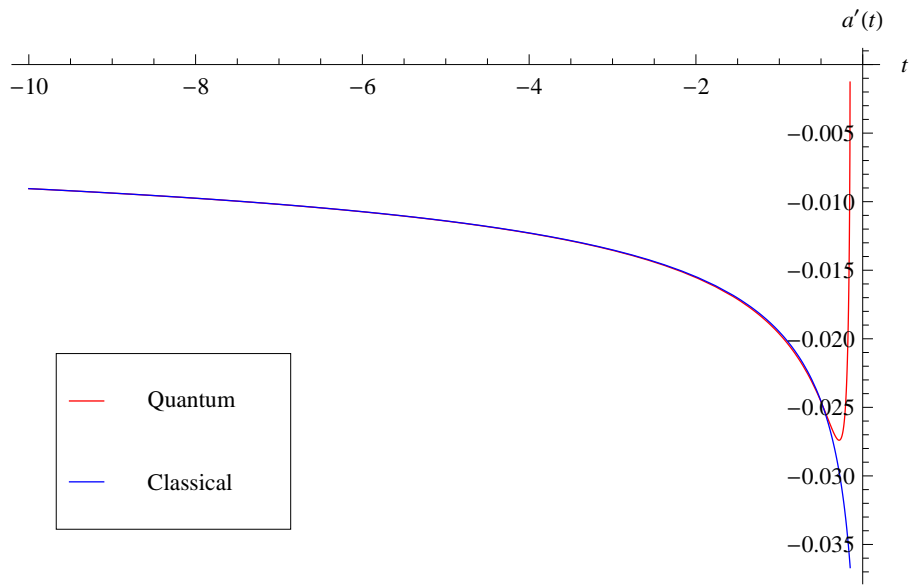}
}
\caption{Collapsing dust-filled universe with $\mu_R=1$ and a single scalar field. The time derivative of the scale factor quickly stops diverging when the quantum correction
becomes active. }
\label{C1}
\end{figure}

Our procedures are similar to those of the previous section. We input the classical solution into the non-local functions. For scalar fields in the case of collapsing dust, this results in
\begin{align}
a \dot{a}^2 - \frac{N_S}{2430 \pi} \left(\frac{19 C_1(t)}{t^2_0 t} + \frac{26 C_2(t)}{t^2_0} \right) = \frac{8 \pi \rho_0}{3}\, .
\end{align}

\noindent The collapse functions are defined as
\begin{align}
C_1(t) = \frac{\log(-\mu_R t)}{t}, \quad C_2(t) = \frac{\log(-\mu_R t) + 1}{t^2}\, .
\end{align}
We note that the initial time in this case is taken to be $-\infty$ as there is no need to cut off the non-local integrals. The normalization time $t_0$ can be chosen arbitrarily but in a regime where the classical behavior remains dominant.

As an example of what happens in a collapsing phase, consider the case $N_S=1$, $\mu_R=1$, shown in Fig. \ref{C1}. Here we see that $\dot{a}(t)$, which is diverging classically, slows down and in fact turns around. This appears as a bouncing solution rather than a singular one. Because of the choice $\mu_R=1$, $\log \mu_R =0$ and there is no local effect in these units.

%%%%%%%%%%%%%%%%%%%%%%%%%%%%%%

\begin{figure*}
\centering
\begin{tabular}{ccc}
\epsfig {file=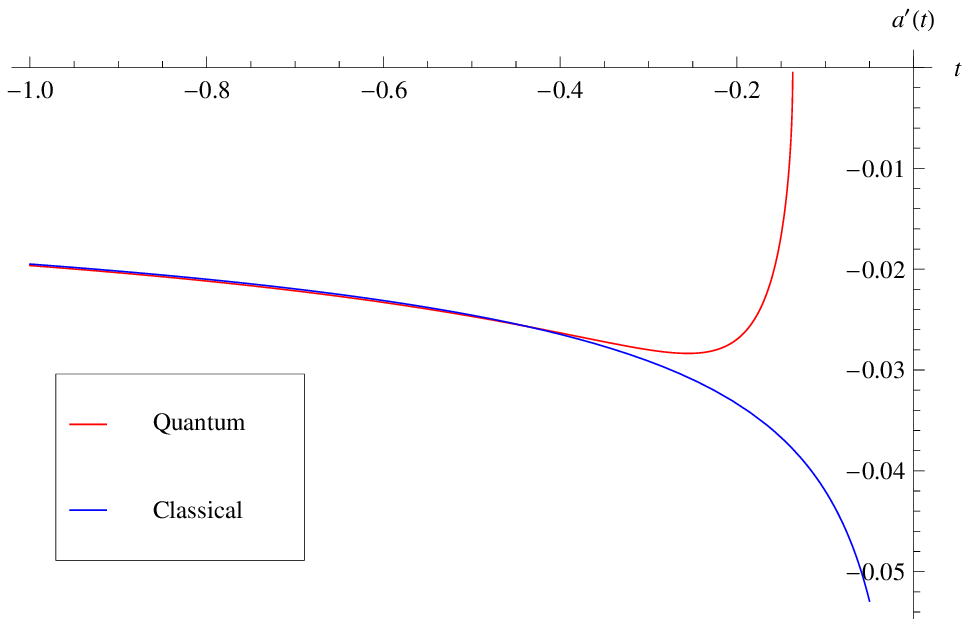,width=0.3\linewidth,clip=}&
\epsfig {file=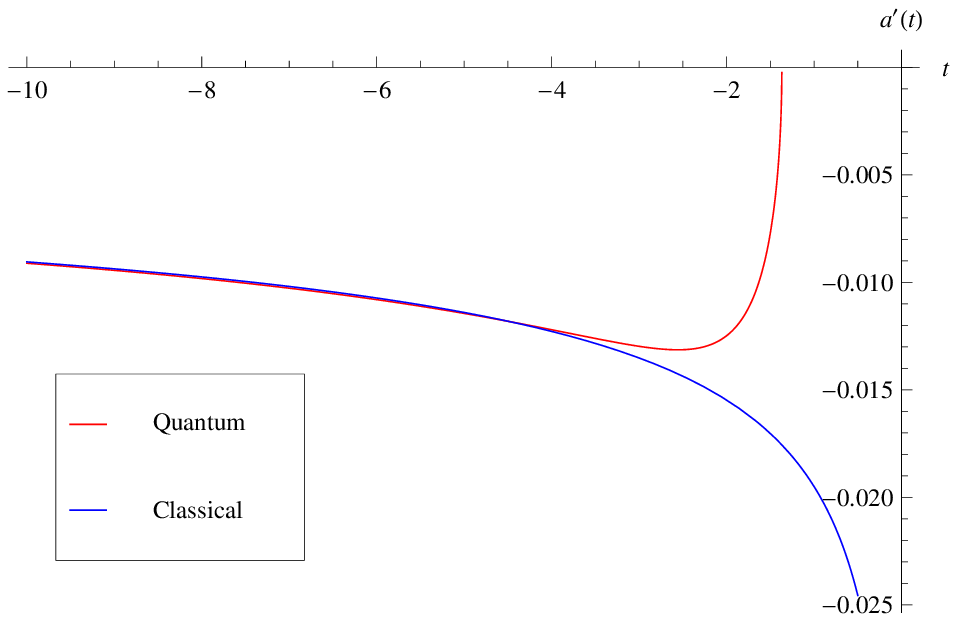,width=0.3\linewidth,clip=}&
\epsfig {file=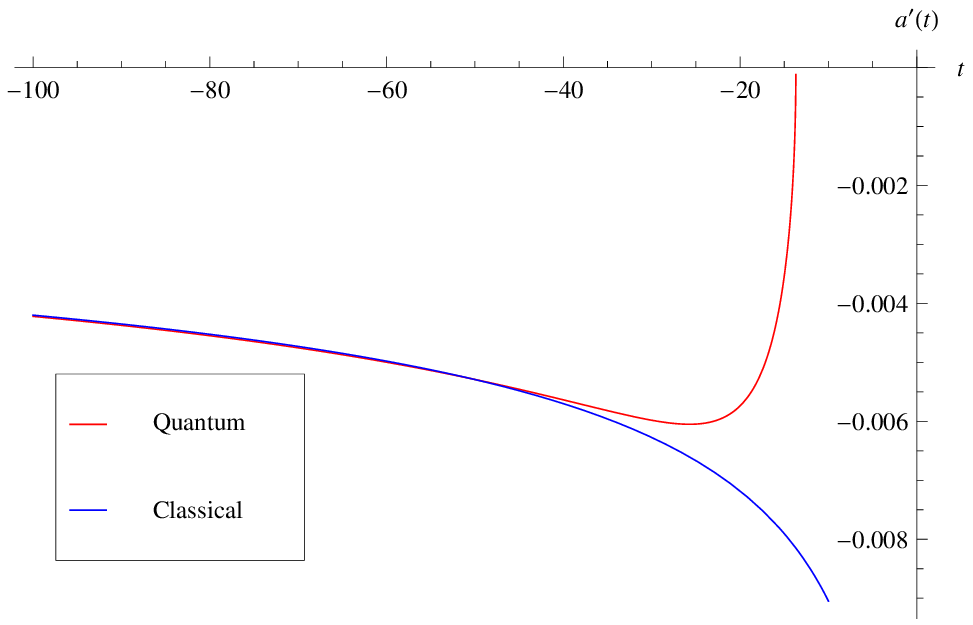,width=0.3\linewidth,clip=}
\end{tabular}
\caption{Varying both the scale $\mu$ and the number of scalar particles $N_S$ in a collapsing dust-filled universe. The plots from left to right involve ($N_S=1,~\mu_R=1$), ($N_S=10^2,~\mu_R=0.1$) and ($N_S=10^4,~\mu_R=0.01$). Note the change of scale along the time axis in the figures. The results illustrate the similarity of the quantum corrections with an energy scale that scales as $E\sim M_P/\sqrt{N}$.}
\label{CmuN}
\end{figure*}

\begin{figure*}
\centering
\begin{tabular}{cc}
\epsfig {file=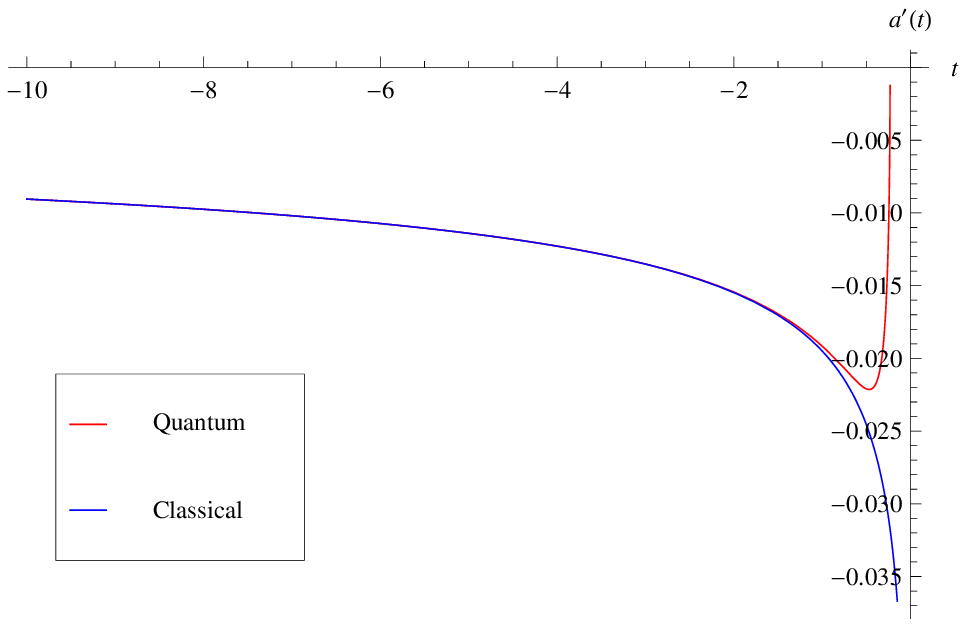,width=0.5\linewidth,clip=}&
\epsfig {file=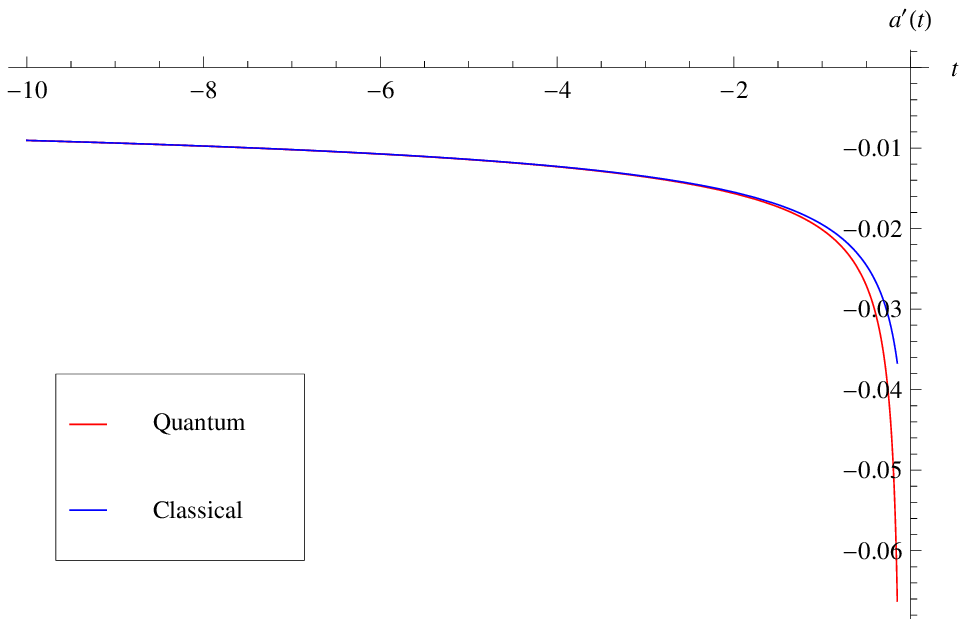,width=0.5\linewidth,clip=}
\end{tabular}
\caption{Varying the scale $\mu_R$ in a collapsing dust-filled universe, with $\mu_R=0.1$ on the left and $\mu_R=10$ on the right.}
\label{Cmu}
\end{figure*}

\begin{figure*}
\centering
\begin{tabular}{cc}
\epsfig {file=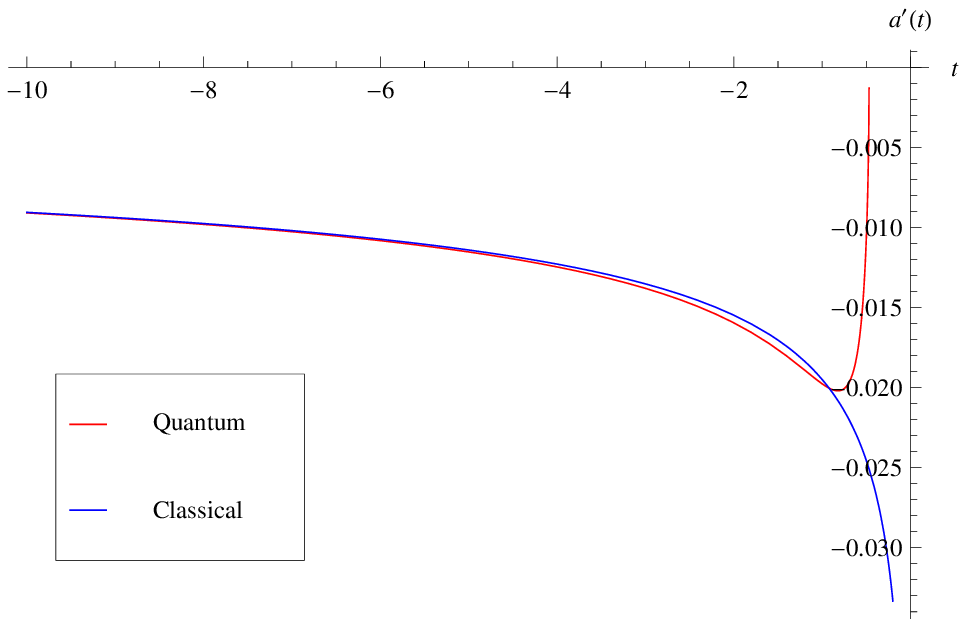,width=0.5\linewidth,clip=}&
\epsfig {file=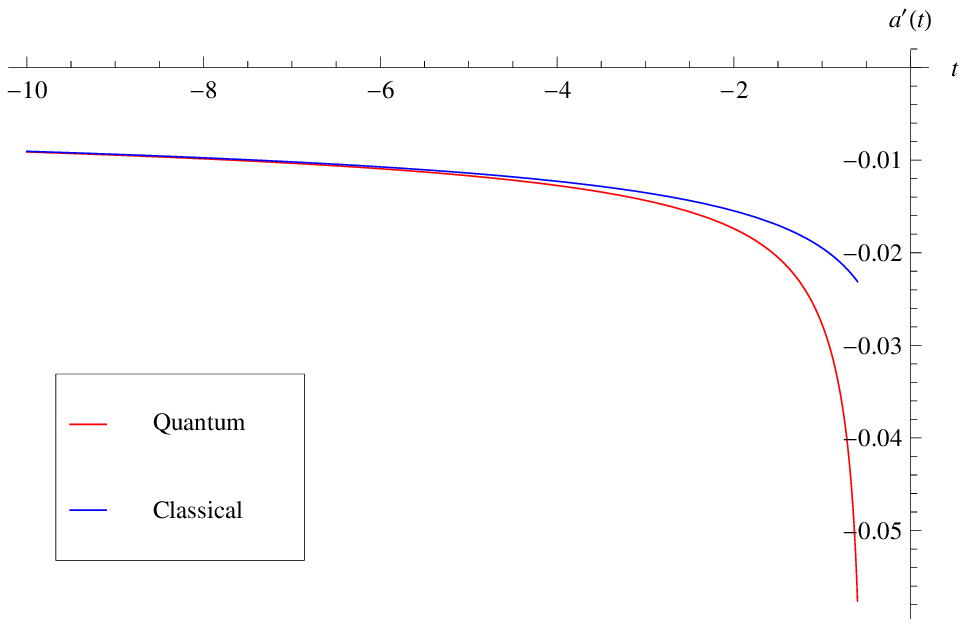,width=0.5\linewidth,clip=}
\end{tabular}
\caption{The effect of graviton loops on a dust-filled universe. These have $\mu_R=0.1$ on the left and $\mu_R=1$ on the right.}
\label{Cgraviton}
\end{figure*}

%%%%%%%%%%%%%%%%%%%%%%%%%
\begin{figure}[ht]
\centerline{
\includegraphics[width=0.7\textwidth]{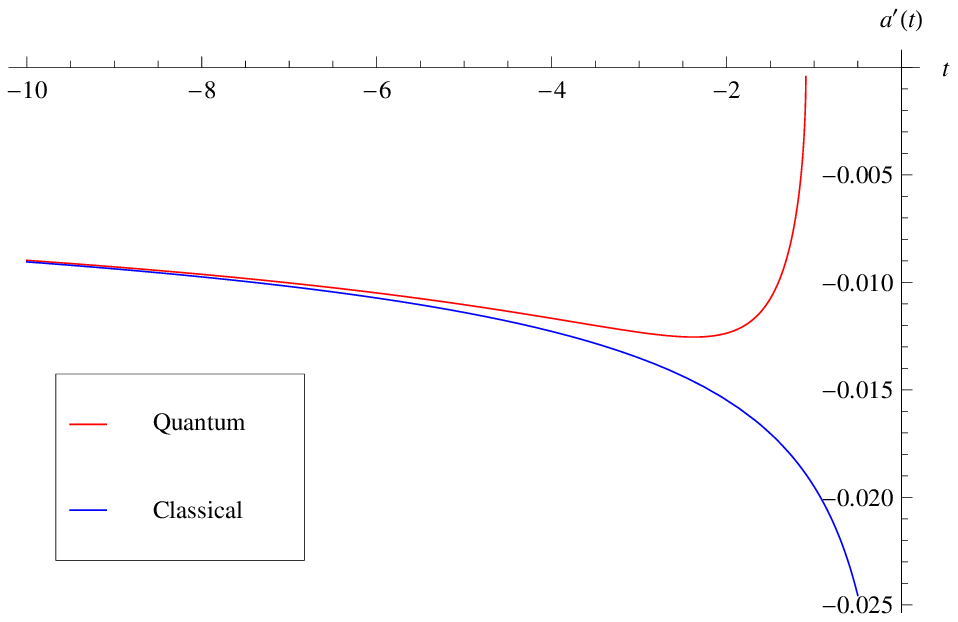}
}
\caption{Collapsing dust-filled universe with the Standard Model particles and a conformally coupled Higgs. The result is purely non-local and hence independent of any scale $\mu_R$. }
\label{CCSM}
\end{figure}
%%%%%%%%%%%%%%%%%%%%%%%%%%%%

If we change the number of scalars, we can lower the energy that this behavior occurs at, in accord with the expected $N$ scaling. This is shown in Fig. \ref{CmuN} by adjusting $N_S$ and $\mu_R$ together such that the number of scalars changes by a factor of 100 between firgures, while $\mu_R$ changes by a factor of 10. This modifies the location of the bounce in a predictable way. The figures look similar even though the horizontal scale changes by a factor of 10 between pictures. The physics does scale as $1/\sqrt{N_S}$ as long as we rescale $\mu_R$ by this factor, and we can have this effect occur well below the Planck scale if the number of scalars is large enough.

However, not all cases lead to singularity avoidance. There is a dependence on the scale $\mu_R$ and for some choices the local terms overwhelm the effect of the non-local terms. This can be seen in Fig. \ref{Cmu}. Here the local terms drive the scale factor in a more singular direction and the singularity happens more rapidly.

%%%%%%%%%%%%%%%%%%%%%%%%
\begin{figure}[ht]
\centerline{
\includegraphics[width=0.7\textwidth]{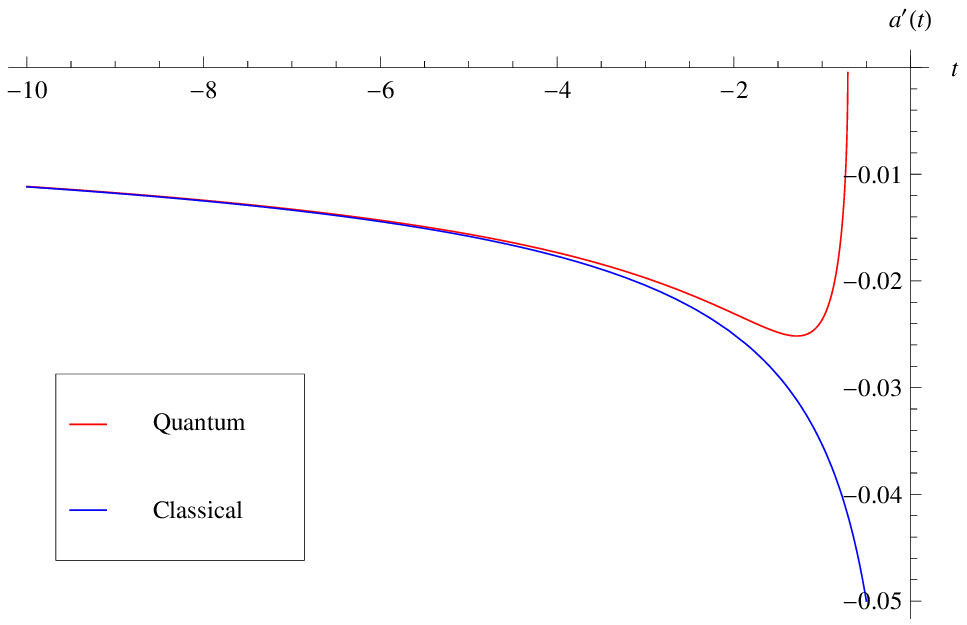}
}
\caption{Collapsing radiation-filled universe with gravitons only considered.}
\label{rgravity}
\end{figure}
%%%%%%%%%%%%%%%%%%%%%%%%%

The bounce is also seen in the case of pure gravity, Fig. \ref{Cgraviton}. The non-local coefficients for the graviton are larger than those for a single
scalar and the change in the scale factor happens at a slightly earlier time than the single scalar case.

%%%%%%%%%%%%%%%%%%%%%%%%%
\begin{figure}[ht]
\centerline{
\includegraphics[width=0.7\textwidth]{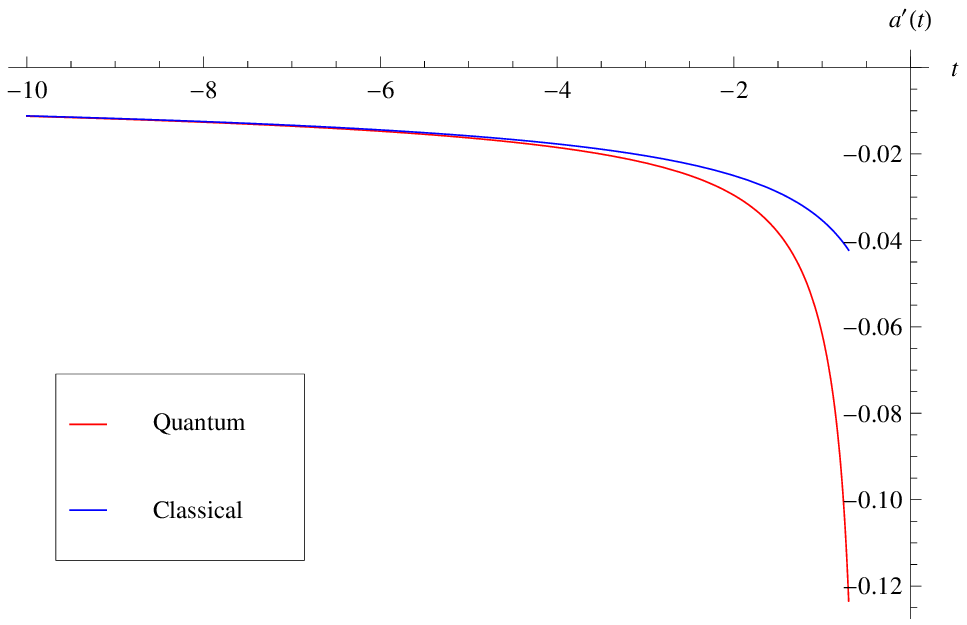}
}
\caption{Collapsing radiation-filled universe all the Standard Model particles included, as well as graviton loops. }
\label{rSMG}
\end{figure}
%%%%%%%%%%%%%%%%%%%%%%%%%

A very interesting case is the Standard Model with a conformally coupled Higgs. As explained in Sec. 5, this situation is purely non-local and completely independent of the parameter $\mu_R$ because in the basis of Eq. \ref{GB}, only the Gauss-Bonnet non-local term contributes and this has no local effect. So this prediction is particularly simple and beautiful. The result with all the Standard Model particles is shown in Fig. \ref{CCSM} and demonstrates the non-local bounce effect in a parameter independent fashion. Note that all conformally coupled fields contribute with the same sign, so that increasing the number of matter fields will
always enhance this effect\footnote{Gravitons are not conformally coupled, but we have checked that their quantum effect (with $\mu$ near unity) is smaller than the effect of the Standard Model particles, and do not change the character of Fig. \ref{CCSM}.}.

For a radiation-filled universe, the effect is always independent of the scale $\mu_R$. With just graviton loops, we see a very similar bounce, see Fig. \ref{rgravity}.
Unfortunately, matter fields have an effect in the opposite direction, and overwhelm the effects of gravity. So with the full set of Standard Model particles plus gravitons, the net effect does not lead to singularity avoidance, as shown in Fig. \ref{rSMG}.

\section{Summary}

Quantum loops bring a unique feature to cosmology, i.e. non-locality. The local classical theory is supplemented by effects which depend on the past behavior of the scale factor. Because of the power-counting theorems of general relativity, these effects are small except at times of large curvature. However, with enough light fields they can become important below the Planck scale.

We have explored the non-local effects that correspond most closely to the graviton vacuum polarization. Our work has been perturbative, in that we treat the new non-local effects to first order only. This is appropriate for a correction that has been calculated at one-loop order only. Actually the large N case can be used to argue that the one-loop result is the most important in the limit of large N. The one-loop integral is proportional to $GN$. For matter fields that have only gravitational interactions, higher loops would either involve extra gravitons in loops (which do not bring extra factors of $N$) or would be the iteration of the simple vacuum polarization. Counting the powers of $G$ and $N$ reveals that the iteration of the one-loop diagram is the only effect of order $(GN)^n$, with other diagrams suppressed by at least a power of $N$.

In addition to the unavoidable use of perturbation theory, we have also approximated the non-local function by its free field behavior. The use of the full propagators is not realistically tractable in a general FLRW space time. The approximation amounts to neglecting higher powers of the curvature which appear in the propagators. This is reasonable when paired with the general use of perturbation theory. The approximation should be good in the region where the non-local integrand, $1/(t-t')$, is the largest. We have not seen any problematic effects from the long-time tail of this integrand.

The most interesting effect uncovered is the tendency towards singularity avoidance in collapsing FLRW universes. The classical theory, with only the Einstein action, collapses towards an inevitable singularity. The quantum effects can oppose this collapse and can turn around converging geodesics. Because of the perturbative treatment, we cannot be certain of the ultimate fate of such effect, but within the limits of our approximations it appears to have the characteristics of a bounce.

There is clearly much more work needed to fully understand the effects of quantum non-locality in general relativity. We will continue our exploration in future work.

\section*{Acknowledgements}

We would like to thank Mohamed Anber for discussions related to the corrections with large number of matter fields. Also B.K.El-M. thanks Amir Azadi for discussions. This work has been supported in part by the US National Science Foundation grant PHY-1205986.

\section*{Appendix: Some aspects of the in-in formalism}

The aim of the in-in formalism is to derive an expression for the time-dependent expectation value of a Heisenberg operator $\mathcal{O}_H(t)$. For systems out of equilibruim, the Hamiltonian has explicit time dependence. For systems under equilibrium, the common practice in perturbation theory is to switch to the interaction picture by splitting the Hamiltonian into free and interaction pieces. For our case, we switch to the interaction picture by splitting the full Hamiltonian to a time-independent piece, which might itself contain interactions, and a time-dependent interaction; $H(t) = H_0 + H_{int}(t)$. Hence,
\begin{align}
\nonumber
\mathcal{O}_H(t) = \mathcal{U}^{\dagger}(t,0) e^{-iH_0t} \, \mathcal{O}_I(t)\, e^{iH_0t} \mathcal{U}(t,0) \equiv S^{\dagger}(t,0) \mathcal{O}_I(t) S(t,0)
\end{align}

\noindent where $\mathcal{U}(t,t^{\prime})$ is the {\em fundamental} time-evolution opertaor and we choose all pictures to coincide at $t=0$. The operator $S(t,t^{\prime})$ is readily seen to satisfy a Schrodinger-like equation whose solution reads
\begin{align}
S(t,t^{\prime}) = T \exp{\left(-i\int_{t^{\prime}}^t dt_1 H_I(t_1)\right)}, \quad H_I(t) \equiv e^{i H_0 t} H_{int}(t) e^{-i H_0 t}\, .
\end{align}

It remains to relate the states in different pictures where it is convenient for our problem to change the reference time such that all pictures coincide at $t=-\infty$. Hence,
\begin{align}
| \Phi \rangle_H = | \Phi(-\infty) \rangle_I\, .
\end{align}

Using the fundamental unitarity property of the time evolution operator, we find the time-dependent expectation value of an arbitrary operator
\begin{align}\label{exval}
\langle \mathcal{O}_H(t) \rangle &= ~  ~_I\langle \Phi(-\infty) | S^{\dagger}(t,-\infty) \mathcal{O}_I(t) S(t,-\infty) |\Phi(-\infty) \rangle_I\, .
\end{align}

As mentioned in the text, it is very useful to insert the identity operator in the form $S^{\dagger}(\infty,t)\, S(\infty,t) = 1$ to the left of the operator
\begin{align}
\langle \mathcal{O}(t) \rangle &=~ ~_I \langle \Phi(-\infty) |_I S^{\dagger}(\infty,-\infty) T\left[\mathcal{O}_I(t) S(\infty,-\infty)\right] |\Phi(-\infty) \rangle_I\, .
\end{align}
One then obtains various propagators - the normal Feynman propagators associated with purely time-ordered contractions, and others associated with mixed contractions. Wick's theorem must then be generalized to include the anti-time-ordered products of fields, which we now describe.

The goal is to modify Wick's theorem to incorporate an anti-time-ordered product of operators. Here, we do not prove the modified theorem but rather only derive the needed expression for our calculation which is
\begin{align}\label{desired}
\widehat{T}[AB]T[CD] = N [ABCD + AB \acontraction{}{C}{}{D}CD + CD \bcontraction{}{A}{}{B}AB + \acontraction{}{C}{}{D}CD \bcontraction{}{A}{}{B}AB + BC \underline{AD} + BD \underline{AC} + AD \underline{BC}+ AC \underline{BD} + \underline{BC}\,\, \underline{AD} + \underline{BD} \,\, \underline{AC}]\, .
\end{align}

\noindent Here, the operators $A,B,C,D$ may represent different fields or the same field evaluated at different spacetime points and the hat denotes the anti-time-ordering symbol. The underline symbol denotes the positive-frequency Wightman function defined in section 3. We also have the usual Feynmann and Dyson propagators
\begin{align}
\acontraction{}{A}{}{B}AB \equiv \langle 0 | T[AB] | 0 \rangle, \quad \bcontraction{}{A}{}{B}AB \equiv \langle 0 | \widehat{T}[AB] | 0 \rangle\, .
\end{align}

\noindent To derive Eqn. (\ref{desired}), we start with the simpler product
\begin{align}\label{3op}
\widehat{T}[AB]C = N [ABC + C \bcontraction{}{A}{}{B}AB + B \underline{AC} + A \underline{BC}]
\end{align}

\noindent which is proved by employing
\begin{align}\label{fund}
\widehat{T}[AB] = N[AB] + \bcontraction{}{A}{}{B}AB, \quad N[AB]C = N[ABC + A \, \underline{BC} + B \, \underline{AC}]\, .
\end{align}

\noindent Left-multiplying Eqn. (\ref{3op}) by an operator, one finds
\begin{align}\label{4op}
\widehat{T}[AB]CD = N[ABCD + AB \, \underline{CD} + AC \, \underline{BD} + BC \, \underline{AD} + CD \bcontraction{}{A}{}{B}AB + BD \, \underline{AC} + AD \, \underline{BC} +\underline{CD} \bcontraction{}{A}{}{B}AB + \underline{BD} \, \, \underline{AC} + \underline{AD}\, \, \underline{BC}]\, .
\end{align}

\noindent The above expression is obtained by deriving the analog of the second equation in (\ref{fund}), albeit with an extra operator to the left. Using the basic defintion of time-ordered products along with Eqn. (\ref{4op}) readily yields Eqn. (\ref{desired}).

\end{document}